\documentclass[a4paper,11pt]{article}
\usepackage[latin1]{inputenc}
\usepackage{graphicx,color,rotating}
\usepackage{amsmath}
\usepackage{natbib}
\usepackage{hyperref}
\usepackage{float}
\usepackage{multirow}
\usepackage[margin=1in]{geometry}
\usepackage{authblk}
\usepackage{setspace}
\usepackage{tikz}
\usetikzlibrary{bayesnet}
\usepackage[normalem]{ulem}
\usepackage{hhline}

\newcommand{\corr}{\mathrm{Corr}}
\newcommand{\dsep}{\mathrm{dsep}}
\newcommand{\es}{$\emptyset$}
\newcommand{\indep}{{\perp\!\!\!\perp}}
\newcommand{\la}{\leftarrow}
\newcommand{\matr}[1]{\boldsymbol{#1}}
\newcommand{\pr}{\mathrm{p}}
\newcommand{\ra}{\rightarrow}
\newcommand{\tr}{\mathrm{tr}}
\newcommand{\transp}{^{\sf t}}
\newcommand{\vect}[1]{\boldsymbol{#1}}

\newcommand{\gcite}{\citep}
\newcommand{\gcitet}{\citet}
\newcommand{\gcitep}{\citealp}

\renewcommand{\vec}{\mathrm{VEC}}
\newcommand{\pfc}{\mathrm{PFC}}
\newcommand{\sma}{\mathrm{SMA}}
\newcommand{\ifg}{\mathrm{IFG}}
\newcommand{\ipl}{\mathrm{IPL}}

\newcommand{\TPM}{theoretically preferred model}
\newcommand{\BFM}{best fit model}

\graphicspath{{Files}}

\title{Multilevel testing of constraints induced by structural equation modeling in fMRI effective connectivity analysis: A proof of concept}

\author{Guillaume Marrelec\textsuperscript{1,*} and Alain Giron\textsuperscript{1,*}\\
\footnotesize \textsuperscript{1} Sorbonne Universit{\'e}, CNRS, INSERM, Laboratoire d'imagerie biom{\'e}dicale, LIB, F-75006, Paris, France\\
\footnotesize \textsuperscript{*} email: firstname.lastname@inserm.fr}

\date{}

\begin{document}

\maketitle

\begin{abstract}
 In functional MRI (fMRI), effective connectivity analysis aims at inferring the causal influences that brain regions exert on one another. A common method for this type of analysis is structural equation modeling (SEM). We here propose a novel method to test the validity of a given model of structural equation. Given a structural model in the form of a directed graph, the method extracts the set of all constraints of conditional independence induced by the absence of links between pairs of regions in the model and tests for their validity in a Bayesian framework, either individually (constraint by constraint), jointly (e.g., by gathering all constraints associated with a given missing link), or globally (i.e., all constraints associated with the structural model). This approach has two main advantages. First, it only tests what is testable from observational data and does allow for false causal interpretation. Second, it makes it possible to test each constraint (or group of constraints) separately and, therefore, quantify in what measure each constraint (or, e..g., missing link) is respected in the data. We validate our approach using a simulation study and illustrate its potential benefits through the reanalysis of published data.
\par
\
\end{abstract}

\section{Introduction} \label{sec:introduction}

A key concept in investigating functional brain interactions in blood oxygen level dependent (BOLD) functional magnetic resonance imaging (fMRI) is effective connectivity, which has been defined as the influence that regions exert on one another \gcite{Friston-1994b}. There are various methods to deal with effective connectivity in fMRI, including, but not restricted to, structural equation modeling (SEM) \gcite{McIntosh-1994, Gonzalez-Lima-1995, Buchel-1997, Bullmore-2000, James-2009, Ramsey_JD-2010}, dynamic causal modeling \gcite{Friston-2003b, Stephan-2007, Daunizeau-2011, Lohmann-2012, Friston-2017} and Granger causality \gcite{Goebel-2003, Roebroeck-2005, Hu-2011, Barrett-2013, Davey-2013}; for partial reviews, see, e.g., \gcitet{Valdes-Sosa-2011, Stephan-2012b, Bielczyk-2019, Jovellar-2019}. We here focus on SEM \gcite{Bollen-1989, Hoyle-2012, Civelek-2018, Davvetas-2020}, which relies on expressing the time course $y_i ( t )$ of one region $i$ as a linear function of the time course of other regions,
\begin{equation} \label{eq:sem:def}
 y_i ( t ) = \sum_{j \neq i} \lambda_{ij} y_q ( t ) + e_i ( t ).
\end{equation}
The model parameters $( \lambda_{ij} )$ are then estimated by optimization of a global cost function that usually depends on the data through the sample covariance matrix only \gcite{Cudeck-1993, Bullmore-2000}. While of great practical interest, blind extraction of the model's structure from data quickly becomes intractable as the number of regions increases. Methods have been proposed that provide heuristics to explore the set of potential models and provides a ``best fit'' according to some (also global) metric \gcite{Bullmore-2000, Ramsey_JD-2010}.
\par
There are two main issues with most methods that have been proposed so far. First, they put the emphasis on links (arrows). By contrast, given observational data, a structural model can only be defined in terms of its constraints of conditional independence, which characterize the \emph{missing} links. It is thus not the links themselves but these missing links that are characteristic of a structural model in the face of data. As a consequence, it can happen that different models (i.e., with different sets of links) generate the same sets of constraints of conditional independence. These models, called observationally equivalent (\gcitep{Stelzl_I-1986}; \gcitep{Verma-1990b}; \gcitep{Mayekawa-1994}; \gcitep[Section~8.11]{Shipley-2000}; \gcitep[Section~5.2.3]{Pearl-2001}), have identical values of data fit regardless of the measure and, therefore, cannot be distinguished from data alone. This means that methods dealing with links do not compare models but classes of (observationally equivalent) models. Application of the main methods of structural equation modeling without being aware of this major feature might give the neuroscientist the false impression that a unique optimal causal model has been found and lead to an erroneous physiological interpretation of such causality.
\par
Another issue is that most methods propose models based on global measures. They therefore base their decision on how a given model globally fits the data. The larger the number of links, the more complex the search space and the smaller the weigh of each link into these global schemes. By contrast, one is often interested in a very specific link, set of links, or pathway and, for instance, the influence of a given task on it. Even in a more general setting, it would still be of interest to be able to test different parts of a model separately, in order to determine where the structural model can be trusted and where it could be improved.
\par
In the present paper, we propose a novel method that avoids the two above-mentioned issues. Given a structural model, it extracts the set of all constraints of conditional independence induced by the absence of links between pairs of regions in the model and tests for their validity in a Bayesian framework, either individually (constraint by constraint), jointly (e.g., by gathering all constraints associated with a given missing link), or even globally (i.e., all constraints associated with a structural model). In other words, we use the relevance of the constraints associated with a structural model as a proxy for the relevance of the structural model itself. With such an approach, observationally equivalent models have the same sets of conditional independence constraints and, therefore, the same set of constraints is tested. Also, it makes it possible to test each constraint (or group of constraints) separately and, therefore, quantify in what measure each constraint (or, e.g., missing link) is respected in the data. This feature gives the neuroscientist a key feedback as to which assumptions underlying the model can be deemed correct and which ones seem to contradict the data. We validate our approach on synthetic data and illustrate its behavior by a reanalysis of published experimental data, showing its potential added value.
\par
We illustrate the approach on a dataset and two structural models originating from \gcitet{Bullmore-2000}. This example has several advantages. It includes a limited number of regions and therefore allows for an exhaustive description of the procedure. Also, it deals with cyclic graphs, which are common in neuroscience/neuroimaging while being quite challenging from a graph-theoretic perspective. Finally, it has already been investigated using other approaches,  including SEM \gcite{Bullmore-2000} and partial correlation \gcite{Marrelec-2007, Marrelec-2009}, giving us a good view of the informative content and limits of the model and data. We use this study to (i) exemplify the extraction of conditional independence constraints from real-life SEMs; (ii) provide a generative model for our synthetic data; and (iii) reanalyze their experimental data.
\par
The outline of the manuscript is the following. In Section~\ref{s:meth}, we introduce and develop the method. In Section~\ref{s:simu}, we assess its validity using a simulation study. In Section~\ref{s:reel}, we reanalyze the experimental data from \gcitet{Bullmore-2000}. Further issues are discussed in Section~\ref{s:disc}.

\section{Method} \label{s:meth}

In this section, we introduce the general frameworks of directed graphs (Section~\ref{ss:meth:dg}) and graphical representation of SEMs (Section~\ref{ss:meth:sem}). We then delve into the extraction of individual conditional independence constraints, first stating the general procedure (Section~\ref{ss:meth:constr}), then applying it to the two structural models of \gcitet{Bullmore-2000} (Section~\ref{ss:meth:ex}). In the particular case of multivariate normal distributions, we show that the contraints have a simple expression in terms of conditional correlation coefficients (Section~\ref{ss:meth:mvd}). Finally, we provide a numerical Bayesian inference procedure to assess the validity of individual constraints and show how it can be used to also test joint and global constraints (Section~\ref{ss:meth:inf}).

\subsection{Directed graphs} \label{ss:meth:dg}

SEMs are often represented in the form of directed graphs and the method that we propose strongly relies on such a representation. In the present section, we quickly introduce the basic notions relative to the theory of directed graphs. For more information, the interested reader can refer to \gcitet{Shipley-2000b} or \gcitet{Pearl-2001d}.
\par
A graph is defined by the set of its vertices (or nodes)---here brain regions---, $\mathcal{R}$, and the set of its edges, or arrows, $\mathcal{A}$. Edges stand for effective connections of the type $i \to j$ from brain region $i$ to brain region $j$. A path from region $i$ to region $j$ is a sequence of regions such that two consecutive regions are connected by an edge. The path is said to be directed if it is possible to go from region $i$ to region $j$ only by following arrows (i.e., from tail to head); it is undirected if the direction of arrows is ignored. $j$ is a descendant of $i$ if there exists a directed path from $i$ to $j$; in that case, we also say that $i$ is an ancestor of $j$. A collider $c$ is a region that has arrows pointing to it, i.e., forming a pattern of the form $i \ra c \la j$. For an application of this terminology to a toy example, see Figure~\ref{fig:ex}.
\par
A directed graph is acyclic if no directed path connects a vertex to itself; otherwise it is cyclic. The theoretical properties of acyclic graphs are much better understood than those of cyclic graphs. However, models originating from neuroscience are quite often cyclic, so both types of graphs need to be taken into account.
\par
A key concept in directed graphs is $d$-separation \gcite{Pearl-1988}, which itself relies on the notion of path blocking. A path between regions $i$ and $j$ is blocked given a set of regions $\mathcal{S}$ if there is a region $k$ on the path for which one of two conditions holds:
\begin{itemize}
 \item $k$ is a non-collider (i.e., $\ra k \ra$ or $\la k \la$) and $k$ is in $\mathcal{S}$;
 \item $k$ is a collider (i.e., $\ra k \la$) and neither $k$ nor any of its descendants is in $\mathcal{S}$.
\end{itemize}
Two regions $i$ and $j$ are then said to be $d$-separated given a set of regions $\mathcal{S}$, denoted $\dsep ( i, j | \mathcal{S} )$, if and only if all paths between regions $i$ and $j$ are blocked by $\mathcal{S}$. $d$-separation is a formal way of describing whether $\mathcal{S}$ can block the flow of information between regions. An illustration is given in Figure~\ref{fig:ex}.

\begin{figure}[!htb]
  \centering
  \begin{tabular}{cc}
    \tikz{ %
     \node[latent] (1) {$1$}; %
     \node[latent, right=1cm of 1, yshift=1cm] (2) {$2$}; %
     \node[latent, right=1cm of 1, yshift=-1cm] (3) {$3$}; %
     \node[latent, right=1cm of 2, yshift=-1cm] (4) {$4$}; %
     \node[latent, right=1cm of 4] (5) {$5$}; %
     \edge[->] {1} {2}; %
     \edge[->] {1} {3};
     \edge[->] {2} {4};
     \edge[->] {3} {4};
     \edge[->] {4} {5};
    } \\
    \ \\
    $\footnotesize \left\{ \begin{array}{ccccccc}
      y_1 ( t ) & = & & & & & e_1 ( t ) \\
      y_2 ( t ) & = & \lambda_{21} y_1 ( t ) & & & + & e_2 ( t ) \\
      y_3 ( t ) & = & \lambda_{31} y_1 ( t ) & & & + & e_3 ( t ) \\
      y_4 ( t ) & = & \lambda_{42} y_2 ( t  ) & + & \lambda_{43} y_3 ( t ) & + & e_4 ( t ) \\
      y_5 ( t ) & = & \lambda_{54} y_4 ( t ) & & & + & e_5 ( t ).
    \end{array} \right.$
  \end{tabular}
  \caption{\textbf{Example of directed graph and SEM.} Top: Directed acyclic graph. In this example, we have $\mathcal{R} = \{ 1, 2, 3, 4, 5, 6 \}$ and $\mathcal{A} = \{ (1, 2), (1, 3), (2, 4), (3, 4), ( 4, 5) \}$. $1 \ra 2 \ra 4 \ra 5$ is a directed path from 1 to 5, $1 \ra 2 \ra 4 \la 3$ an undirected path between 1 and 3. 4 and 5 are descendant of 2. 1, 2, and 3 are ancestors of 4. 4 is a collider, since both 2 and 3 are pointing to it.  The path $1 \ra 2 \ra 4 \ra 5$ is blocked by 2, since 2 is a non-collider which is on the path; it is also blocked by $\{ 2, 3 \}$ for the same reason. However, this path is not blocked by 3 alone, since 3 is a non-collider but is not on the path. The path $2 \ra 4 \la 3$ is blocked by $\emptyset$ (the empty set), since the only node on the path is 4, which is a collider and does not belong to $\emptyset$. However, this path is neither blocked by 4 nor by 5, since 4 is a collider that is on the path and 5 is its descendant. 1 and 4 are $d$-separated by $\{ 2, 3 \}$, since both paths connecting 1 and 4 (i.e., $1 \ra 2 \ra 4$ and $1 \ra 3 \ra 4$) are blocked by $\{ 2, 3 \}$; however, they are not $d$-separated by $2$ only, since this node does not block the path $1 \ra 3 \ra 4$. Inverting the link $1 \ra  2$ yields a graph that is observationally equivalent, as does inverting the link $1 \ra 3$. Bottom: Structural equation model whose graphical representation is given by the top graph.} \label{fig:ex}
\end{figure}

\subsection{Graphical representation of SEMs} \label{ss:meth:sem}

As mentioned earlier, directed graphs are a convenient and powerful way to represent SEMs. Assume that the state of every region $i$ in a network $\mathcal{R}$ of $R$ regions is quantified by a variable $y_i$ and that the global, $R$-dimensional variable $\vect{y} = ( y_1, \dots, y_R )$, describing the state of the whole network, is modeled by a SEM with relationships of the form of Equation~\eqref{eq:sem:def}. Note that $\lambda_{ij}$ reads as the part of the signal of $i$ that is contributed by region $j$. Then $\vect{y}$ can be associated with a directed graph, where each variable $y_i$ is represented by a vertex $i$, and where there is an arrow from region $j$ to node $i$ whenever $\lambda_{ij} \neq 0$.

\subsection{Constraints imposed by a structural model} \label{ss:meth:constr}

It has been shown that for a wide variety of structural models, including acyclic causal models, as well as cyclic causal models in which all variables are discrete or in which functional relationships are linear, each relationship of $d$-separation can be translated into a relationship of conditional independence \gcite[Section~2.8]{Shipley-2000b}. More precisely, every time two regions $i$ and $j$ are $d$-separated by a set of regions $\mathcal{S}$, then $y_i$ and $y_j$ must be conditionally independent given $\vect{y}_{\mathcal{S}}$, denoted
\begin{equation}
 y_i \indep y_j | \vect{y}_{\mathcal{S}}.
\end{equation}
Since each relationship of $d$-separation within an SEM is mirrored by a relationship of conditional independence, extracting the set of all relationships of $d$-separation for a model yields all relationships of conditional independence that must be verified in the data if the model is correct. Note that all these relationships are not necessarily independent from one another, as some statements can be predicted from others. In particular, we expect constraints involving the same pair of variables $( i, j )$ to be highly correlated. We here decide to keep and test all constraints. This question is further discussed in Section~\ref{s:disc}.

\subsection{Example} \label{ss:meth:ex}

We use two structural models from the literature \gcite{Bullmore-2000} to exemplify the extraction of conditional independence constraints from real-life directed graphs.
\par
\gcitet{Bullmore-2000} considered SEM analysis of a task requiring semantic decision and subvocal rehearsal. The following $R = 5$ left hemispheric cortical regions of interest were selected: the ventral extrastriate cortex (VEC), the prefrontal cortex (PFC), the supplementary motor area (SMA), the inferior frontal gyrus (IFG), and the inferior parietal lobule (IPL). Thus, the set of vertices is given by
$$\mathcal{R} = \left\{ \mbox{VEC, PFC, SMA, IFG, IPL} \right \}.$$
In their study, they introduced two structural models. They first proposed a plausible structural model based on anatomical and functional considerations. The resulting model, henceforth referred to as the ``\TPM'' (or ``TP'' model; see Figure~\ref{fig:meth:ex}, left). They also applied a procedure implemented in the LISREL proprietary software package\footnote{http://www.ssicentral.com/lisrel/}, yielding a ``\BFM'', henceforth referred to as such (or ``BF'' model). The resulting model is schematized in Figure~\ref{fig:meth:ex}, right. Note that both structural models are cyclic. Two cycles in the \TPM\ are
\begin{center}
  IPL $\ra$ VEC $\ra$ IPL
\end{center}
 and
\begin{center}
  IPL $\ra$ VEC $\ra$ PFC $\ra$ SMA $\ra$ IFG $\ra$ IPL,
\end{center}
while, for the \BFM\ we have
\begin{center}
  IPL $\ra$ VEC $\ra$ PFC $\ra$ IFG $\ra$ IPL.
\end{center}
and
\begin{center}
  IPL $\ra$ VEC $\ra$ PFC $\ra$ SMA $\ra$ IPL.
\end{center}

\tikzset{region/.style={draw,circle,minimum width=3em,thick,fill=white,font=\small,inner sep=0}}
\tikzset{lien/.style={very thick,->,>=stealth'}}

\begin{figure}[!htbp]
  \centering
  \begin{tabular}{ccc }
   Theoretically preferred (TP) model & \ & Best fit (BF) model \\
   \ \\
   \begin{tikzpicture}[scale=0.8]
    \node[region] (pfc) at (0,0) {PFC};
    \node[region] (sma) at (2,2) {SMA};
    \node[region] (ifg) at (3,-0.7) {IFG};
    \node[region] (ipl) at (5.5,0.7) {IPL};
    \node[region] (vec) at (5.5,-2) {VEC};
    \draw[lien] (pfc) -- (sma);
    \draw[lien] (sma) -- (ifg);
    \draw[lien] (ifg) -- (ipl);
    \draw[lien] (ipl) to[bend right] (vec);
    \draw[lien] (vec) to[bend right] (ipl);
    \draw[lien] (vec.south west) to[bend left=45] (pfc);
   \end{tikzpicture}
   & &
   \begin{tikzpicture}[scale=0.8]
    \node[region] (pfc) at (0,0) {PFC};
    \node[region] (sma) at (2,2) {SMA};
    \node[region] (ifg) at (3,-0.7) {IFG};
    \node[region] (ipl) at (5.5,0.7) {IPL};
    \node[region] (vec) at (5.5,-2) {VEC};
    \draw[lien] (pfc) -- (sma);
    \draw[lien] (pfc) -- (ifg);
    \draw[lien] (sma) -- (ipl);
    \draw[lien] (ifg) -- (ipl);
    \draw[lien] (ipl) --  (vec);
    \draw[lien] (vec.south west) to[bend left=45] (pfc);
   \end{tikzpicture}
  \end{tabular}
  \caption{\textbf{Example of structural models. ``Theoretically preferred'' (TP) model (left) and ``best fit'' (BF) model (right) from \gcitet{Bullmore-2000}.}} \label{fig:meth:ex}
\end{figure}
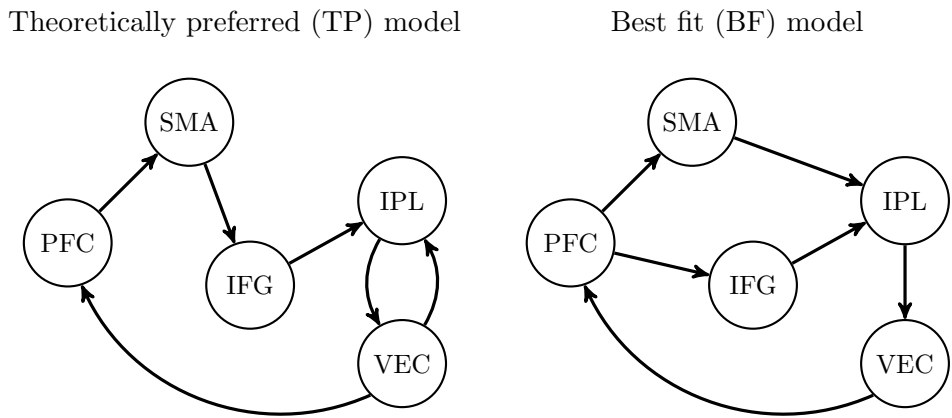

We are now in position to extract the conditional independence constraints from the TP model (Section~\ref{sss:meth:ex:tp}) and the BF model (Section~\ref{sss:meth:ex:bf}).

\subsubsection{Theoretically preferred model} \label{sss:meth:ex:tp}

For the \TPM, the following links are missing: between VEC and SMA; VEC and IFG; PFC and IFG, PFC and IPL, SMA and IPL. Reviewing all paths between these pairs of regions (see Table~\ref{tab:meth:ex}) leads to the following sets of $d$-separation and corresponding constraints on conditional correlation---
\begin{itemize}
 \item Between VEC and SMA: There are three potential paths between both regions. PFC must be in the conditioning set, otherwise VEC $\ra$ PFC $\ra$ SMA is not blocked. IFG must also be in the conditioning set, otherwise VEC $\la$ IPL $\la$ IFG $\la$ SMA is not blocked. Since IFG also blocks VEC $\ra$ IPL $\la$ IFG $\la$ SMA, the set $\{$ PFC, IFG $\}$ is sufficient. Adding IPL in the conditioning set does not change the conclusion. We therefore have the following two constraints:
 $$\begin{array}{cccc}
  C_1: & \corr [ \vec, \sma | \pfc, \ifg ] & = & 0 \\
  C_2: & \corr [ \vec, \sma | \pfc, \ifg, \ipl ] & = & 0.
 \end{array}$$
 \item Between VEC and IFG: IPL must be in the set, otherwise IFG $\ra$ IPL $\ra$ VEC is not blocked. However, IPL activates the path IFG $\ra$ IPL $\la$ VEC, with no possibility to block it with a non-collider. As a consequence, no set of regions can $d$-separate VEC and IFG.
 \item Between PFC and IFG: SMA must be in the conditioning set to block PFC $\ra$ SMA $\ra$ IFG. VEC must also be in the conditioning set to block both PFC $\la$ VEC $\la$ IPL $\la$ IFG and PFC $\la$ VEC $\ra$ IPL $\la$ IFG. Adding IPL keep the conclusion unchanged. We therefore have the following two constraints:
 $$\begin{array}{cccc}
  C_3: & \corr [ \pfc, \ifg | \vec, \sma ] = 0 \\
  C_4: & \corr [ \pfc, \ifg | \vec, \sma, \ipl ] = 0.
 \end{array}$$
 \item Between PFC and IPL: VEC must be in the conditioning set to block both PFC $\la$ VEC $\la$ IPL and PFC $\la$ VEC $\ra$ IPL. As to the first path, it is blocked by either SMA or IFG, or the two. As a consequence, we have three constraints:
 $$\begin{array}{cccc}
  C_5: & \corr [ \pfc, \ipl | \vec, \ifg ] & = & 0 \\
  C_6: & \corr [ \pfc, \ipl | \vec, \sma ] & = & 0 \\
  C_7: & \corr [ \pfc, \ipl | \vec, \sma, \ifg ] & = & 0.
 \end{array}$$
 \item Between SMA and IPL: IFG must be in the conditioning set to block the first path. Both the second and third path are blocked by VEC and/or PFC. We consequently have the following constraints:
 $$\begin{array}{cccc}
  C_8: & \corr [ \sma, \ipl | \pfc, \ifg ] & = & 0 \\
  C_9: & \corr [ \sma, \ipl | \vec, \ifg ] & = & 0 \\
  C_{10}: & \corr [ \sma, \ipl | \vec, \pfc, \ifg ] & = & 0.
 \end{array}$$
\end{itemize}

\subsubsection{Best fit model} \label{sss:meth:ex:bf}

For the \BFM, the following links are missing: between VEC and SMA; VEC and IFG; and PFC and IPL. Reviewing all paths between these pairs of regions (see Table~\ref{tab:meth:ex}) leads to---
\begin{itemize}
 \item Between VEC and SMA: PFC and IPL must be in the conditioning set, otherwise either one of the first two paths connecting these regions is not blocked. This set is sufficient, since both nodes also block the third path. Adding IFG in the conditioning set does not change the conclusion. We therefore have the following two constraints:
 $$\begin{array}{cccc}
  C_1': & \corr [ \vec, \sma | \pfc, \ipl ] & = & 0 \\
  C_2': & \corr [ \vec, \sma | \pfc, \ifg, \ipl ] & = & 0.
 \end{array}$$
 \item Between VEC and IFG: for the same reason as above, PFC and IPL must be in the conditioning set. This set is sufficient, since both nodes also block the third path. Adding SMA in the conditioning set does not change the conclusion. We therefore have the following two constraints:
 $$\begin{array}{cccc}
  C_3': & \corr [ \vec, \sma | \pfc, \ipl ] & = & 0 \\
  C_4': & \corr [ \vec, \sma | \pfc, \sma, \ipl ] & = & 0.
 \end{array}$$
 \item Between PFC and IPL: VEC, SMA, and IFG must all be in the conditioning set to block any of the three paths. As a consequence, we have one constraint:
 $$\begin{array}{cccc}
  C_5': & \corr [ \pfc, \ipl | \vec, \sma, \ifg ] & = & 0.
 \end{array}$$
 \item Between SMA and IFG: PFC must be in the conditioning set to block the first path. This node also blocks the second and fourth path. However, PFC is also a descendant of IPL, which is a collider in the third path. As a consequence, it unblocks the third path, with no possibility to block it, since it has only one intermediary node (which is the collider). As a consequence, no set of regions can $d$-separate SMA and IFG.
\end{itemize}

\begin{table}[!htbp]
 \centering
 \caption{\textbf{Example of structural models.} Review of missing connections and corresponding paths in the theoretically preferred (TP) model and the best fit (BF) model.} \label{tab:meth:ex}
 \begin{tabular}{c|l|l|l|l}
   model & connection & paths & non-collider(s) & collider(s) \\
   \hline
   TP & VEC--SMA & VEC $\ra$ PFC $\ra$ SMA & PFC & \es \\
      &          & VEC $\ra$ IPL $\la$ IFG $\la$ SMA & IFG & IPL \\
      &          & VEC $\la$ IPL $\la$ IFG $\la$ SMA & IPL, IFG & \es \\
   \hhline{~----}
      & VEC--IFG & VEC $\ra$ PFC $\ra$ SMA $\ra$ IFG & PFC, SMA & \es \\
      &          & VEC $\ra$ IPL $\la$ IFG & \es & IPL \\
      &          & VEC $\la$ IPL $\la$ IFG & IPL & \es \\
   \hhline{~----}
      & PFC--IFG & PFC $\ra$ SMA $\ra$ IFG & SMA & \es \\
      &          & PFC $\la$ VEC $\ra$ IPL $\la$ IFG & VEC & IPL \\
      &          & PFC $\la$ VEC $\la$ IPL $\la$ IFG & VEC, IPL & \es \\
   \hhline{~----}
      & PFC--IPL & PFC $\ra$ SMA $\ra$ IFG $\ra$ IPL & SMA, IFG & \es \\
      &          & PFC $\la$ VEC $\ra$ IPL & VEC & \es \\
      &          & PFC $\la$ VEC $\la$ IPL & VEC & \es \\
   \hhline{~----}
      & SMA--IPL & SMA $\ra$ IFG $\ra$ IPL & IFG & \es \\
      &          & SMA $\la$ PFC $\la$ VEC $\ra$ IPL & PFC, VEC & \es \\
      &          & SMA $\la$ PFC $\la$ VEC $\la$ IPL & PFC, VEC & \es \\
   \hline
   \hline
   BF & VEC--SMA & VEC $\ra$ PFC $\ra$ SMA & PFC & \es \\
      &          & VEC $\la$ IPL $\la$ SMA & IPL & \es \\
      &          & VEC $\la$ IPL $\la$ IFG $\la$ PFC $\ra$ SMA & IPL, IFG, PFC & \es \\
   \hhline{~----}
      & VEC--IFG & VEC $\ra$ PFC $\ra$ IFG & PFC & \es \\
      &          & VEC $\la$ IPL $\la$ IFG & IPL & \es \\
      &          & VEC $\la$ IPL $\la$ SMA $\la$ PFC $\ra$ IFG & IPL, SMA, PFC & \es \\
   \hhline{~----}
      & PFC--IPL & PFC $\ra$ SMA $\ra$ IPL & SMA & \es \\
      &          & PFC $\ra$ IFG $\ra$ IPL & IFG & \es \\
      &          & PFC $\la$ VEC $\la$ IPL & VEC & \es \\
   \hhline{~----}
      & SMA--IFG & SMA $\la$ PFC $\ra$ IFG & PFC & \es \\
      &          & SMA $\la$ PFC $\la$ VEC $\la$ IPL $\la$ IFG & PFC, VEC, IPL & \es \\
      &          & SMA $\ra$ IPL $\la$ IFG & \es & IPL \\
      &          & SMA $\ra$ IPL $\ra$ VEC $\ra$ PFC $\ra$ IFG & IPL, VEC, PFC & \es
 \end{tabular}
\end{table}

\subsection{Case of multivariate normal distributions} \label{ss:meth:mvd}

A common assumption in fMRI is that the data follow a multivariate normal distribution. In the framework of SEMs, this is a direct consequence of the assumption that the noise components $e_i(t)$ are Gaussian distributed. If $\vect{y}$ is multivariate normal with covariance matrix $\matr{\Sigma}$, conditional independence is characterized by a zero conditional correlation. Each relationship of $d$-separation $\dsep ( i, j | \mathcal{S} )$ that can be read off the directed graph associated with the SEM can therefore also be associated with one constraint of the form
\begin{equation} \label{eq:constr:mvd}
 \rho_{i,j|\mathcal{S}} = \corr [ y_i, y_j | \vect{y}_{ \mathcal{S} } ] = 0.
\end{equation}
This coefficient of conditional correlation can be obtained from $\matr{\Sigma}$ as follows. First, we discard the part of $\matr{\Sigma}$ not related to $\mathcal{T} = \mathcal{S} \cup \{ i ,j \}$. We then partition the remaining matrix $\matr{\Sigma}_{\mathcal{T}}$ as
$$\begin{pmatrix}
 \matr{\Sigma}_{ \{ i, j \} }  & \matr{\Sigma}_{ \{ i, j \}, \mathcal{S} } \\
 \matr{\Sigma}_{ \mathcal{S}, \{ i, j \} } & \matr{\Sigma}_{ \mathcal{S} }
\end{pmatrix}.$$
$\matr{\Sigma}_{ \{ i, j \} }$ is the part of the covariance matrix that is specific of regions $i$ and $j$, $\matr{\Sigma}_{ \mathcal{S} }$ the part of the covariance matrix that is specific to regions in $\mathcal{S}$, while $\matr{\Sigma}_{ \{ i, j \}, \mathcal{S} }$ contains the covariances between regions in $\mathcal{S}$ on the one hand and, on the other hand, regions $i$ and $j$. We then compute the 2-by-2 conditional covariance matrix of $\{i,j\}$ given $\mathcal{S}$ as \gcite[Section~2.5.1]{Anderson_TW-1958}
$$\matr{\Sigma}_{ \{ i, j \} | \mathcal{S} } = \matr{\Sigma}_{ \{ i, j \} } - \matr{\Sigma}_{ \{ i, j \}, \mathcal{S} } \matr{\Sigma}_{ \mathcal{S} } ^ {-1} \matr{\Sigma}_{ \mathcal{S}, \{ i, j \} },$$
and the corresponding conditional correlation coefficients by normalization of the covariance coefficient,
\begin{equation} \label{eq:corrcond}
 \rho_{i,j|\mathcal{S}} = \frac{ \left( \matr{\Sigma}_{ \{ i, j \} | \mathcal{S} } \right)_{12} } { \sqrt{ \left( \matr{\Sigma}_{ \{ i, j \} | \mathcal{S} } \right)_{11} \left( \matr{\Sigma}_{ \{ i, j \} | \mathcal{S} } \right)_{22} } }.
\end{equation}

\subsection{Inference} \label{ss:meth:inf}

Assume that the SEM leads to the expression of $K$ constraints of the form given by Equation~\eqref{eq:constr:mvd},  each constraint $C_k$ being expressed as
\begin{equation}
 \rho_{ i_k,j_k | \mathcal{S}_k } = 0, \qquad k = 1, \dots, K.
\end{equation}
Starting from a dataset $D = \{ \vect{y}_1, \dots, \vect{y}_N \}$ of $N$ samples assumed to be independent and identically distributed (i.i.d.) realizations of a multivariate normal distribution with unknown mean $\vect{\mu}$ and covariance matrix $\matr{\Sigma}$, the goal of the present section is to propose a way to simultaneously assess the validity of all constraints. To this aim, we use a standard Bayesian analysis complemented with a numerical sampling scheme, as described here. For the sake of simplicity, we denote
\begin{equation}
 \vect{\rho} = ( \rho_{ i_1,j_1 | \mathcal{S}_1 }, \dots, \rho_{ i_K,j_K | \mathcal{S}_K } )
\end{equation}
the set of all conditional correlation coefficients of interest. In a Bayesian analysis, the information brought by the data is summarized by the posterior distribution of the parameters given the data, $\pr ( \vect{\rho} | D )$. While this distribution cannot be expressed in closed form in the present case, it can easily be approximated using a standard result regarding the posterior distribution of the covariance matrix, $\pr ( \matr{\Sigma} | D )$ (Section~\ref{sss:meth:inf:sigma}), together with a numerical sampling scheme (Section~\ref{sss:meth:inf:rho}). From there, hypothesis testing can be performed at the individual, joint, or global level (Section~\ref{sss:meth:inf:test}).

\subsubsection{Posterior distribution of $\matr{\Sigma}$} \label{sss:meth:inf:sigma}

Assuming a noninformative Jeffreys prior for the covariance matrix and letting $\vect{m}$ be the sample average,
$$\vect{m} = \sum_{ n = 1 } ^ N \vect{y}_n,$$
and $\matr{S}$ the sample sum-of-square matrix,
$$\matr{S} = \sum_{ n = 1 } ^ N ( \vect{y}_n - \vect{m} ) ( \vect{y}_n - \vect{m} ) \transp,$$
the posterior distribution of $\matr{\Sigma}$ is inverse Wishart with $N-1$ degrees of freedom and scale matrix $\matr{S}$ \gcite[Section~3.6]{Gelman-1998},
\begin{equation} \label{eq:post}
 \pr ( \matr{\Sigma} | D ) \propto | \matr{\Sigma} | ^ { - \frac{ N + D } { 2 } } \exp \left[ - \frac{ 1 } { 2 } \tr \left( \matr{S} \matr{\Sigma} ^ { - 1 } \right) \right].
\end{equation}

\subsubsection{Posterior distribution of $\vect{\rho}$} \label{sss:meth:inf:rho}

We start by numerically sampling $\matr{\Sigma}$ from its posterior distribution, Equation~\eqref{eq:post}, which can easily be done \gcite[Appendix~A]{Gelman-1998}. From each sample $\matr{\Sigma}^{[l]}$, $l=1,\dots,L$ (e.g., $L = 10^5$) and each constraint $C_k$, one can then compute the conditional correlation coefficient $\rho_{ i_k,j_k | \mathcal{S}_k }^{[l]}$ associated with $C_k$ using the procedure detailed in Section~\ref{ss:meth:mvd} and leading to Equation~\eqref{eq:corrcond}. This procedure yields a sample from the posterior distribution of interest, $\pr ( \vect{\rho} | D )$.

\subsubsection{Hypothesis testing} \label{sss:meth:inf:test}

We are now in position to provide a general test that uses the numerical sample from $\pr ( \vect{\rho} | D )$ to test all (i.e., individual,  joint, and global) constraints induced by a model in a similar fashion. This test is based on a (multivariate) normal approximation of the distributions of interest. A (multivariate) normal distribution is fully characterized by its mean and variance. In our case, approximations for the posterior mean and covariance matrix can readily be computed using their sample counterparts. Let $\tilde{\vect{\rho}}$ be the $\tilde{K}$-dimensional ($1 \leq \tilde{K} \leq K$) subvector of $\vect{\rho}$ that we wish to test, and $\vect{c}$, respectively $\matr{V}$, be the sample mean, respectively variance (or covariance matrix), of the numerical sample $( \tilde{\vect{\rho}} ^ {[l]} )$ approximating $\pr ( \tilde{\vect{\rho}} | D )$. We then define deviance as
\begin{equation} \label{eq:def:dev}
 d ( \tilde{\vect{\rho}} ) = ( \tilde{\vect{\rho}} - \vect{c} ) \transp \matr{V} ^ { - 1 } ( \tilde{\vect{\rho}} - \vect{c} ).
\end{equation}
It is a squared Mahalanobis distance which characterizes how $\tilde{\vect{\rho}}$ is close to $\vect{c}$. If $\tilde{\vect{\rho}}$ were normal distributed, then $d ( \tilde{\vect{\rho}} )$ would provide the contours of constant probability for $\tilde{\vect{\rho}}$ and would be chi-square distributed with $\tilde{K}$ degrees of freedom. 
\par
We finally define $p$ as the probability that $d ( \tilde{\vect{\rho}} ) < d ( \vect{0} )$. $p$ is a measure of how plausible the null hypothesis $\tilde{\vect{\rho}} = \vect{0}$ is \gcite{Tanner-1994, Kershaw-1999, Marrelec-2003b}. This probability could be computed using the chi-square approximation. Here, we rather rely on the numerical sample and approximate $p$ as the fraction of samples for which $d ( \tilde{\vect{\rho}} ^ { [ l ] } )$ is smaller than $d ( \vect{0} )$,
\begin{equation}
 p \approx \frac{1}{L} \# \left\{ l: \, d \left( \tilde{\vect{\rho}} ^ { [ l ] } \right) < d ( \vect{0} ) \right\}.
\end{equation}
Finally, if a decision is required, a significance level $\alpha$ can be set (e.g., $\alpha = 0.05$). All tests such that $p < \alpha$ are declared significant, and the corresponding null hypotheses are rejected.

\section{Simulation study} \label{s:simu}

We assessed the validity of our approach by applying it to synthetic data. The data generation process, the analysis, the evaluation scheme, and the main results are presented in Sections \ref{ss:simu:data}, \ref{ss:simu:anal}, \ref{ss:simu:eval}, and \ref{ss:simu:res}, respectively.

\subsection{Data generation} \label{ss:simu:data}

We generated synthetic data using the TP and BF models introduced earlier (see Section~\ref{ss:meth:ex}). For the model parameters (path coefficients and noise variances), we used the values that were inferred by \gcitet{Bullmore-2000} from their data and are summarized in our Table~\ref{tab:simu:coef}. For each model, we generated 1000 sample covariance matrices, for a total of $1000$ (samples per structural model) $\times 2$ (models)  $= 2000$ sample covariance matrices.

\begin{table}[!hbt]
  \centering
  \caption{\textbf{Simulation study.} Parameter values used for data generation: path coefficients for the theoretically preferred model (top) and the best fit model (middle), as well as residual variance (bottom) (values from \gcitep{Bullmore-2000}).} \label{tab:simu:coef}
  \begin{tabular}{c|ccccc}
    $\lambda_{\mathrm{TP}}$ & VEC   & PFC   & SMA   & IFG   & IPL \\
    \hline
    VEC &  0    & 0    & 0    & 0    & 0.80 \\
    PFC &  0.59 & 0    & 0    & 0    & 0    \\
    SMA &  0    & 0.60 & 0    & 0    & 0    \\
    IFG &  0    & 0    & 0.31 & 0    & 0    \\
    IPL & -0.16 & 0    & 0    & 0.52 & 0    \\
    \\
    $\lambda_{\mathrm{BF}}$ & VEC   & PFC   & SMA   & IFG   & IPL \\
    \hline
    VEC &  0    & 0    & 0    & 0    & 0.61 \\
    SMA &  0    & 0.58 & 0    & 0    & 0    \\
    IFG &  0    & 0.43 & 0    & 0    & 0    \\
    IPL &  0    & 0    & 0.27 & 0.58 & 0    \\
    \\
    $\psi$ & 0.825 & 0.868 & 0.870 & 0.881 & 0.851
  \end{tabular}
\end{table}

\subsection{Analysis} \label{ss:simu:anal}

For each sample covariance matrix, we tested the relevance of the constraints originating from both the \TPM\ and the \BFM. More specifically, for each sample covariance matrix and model, we tested the validity of the following constraints:
\begin{itemize}
 \item For the \TPM: 10 individual constraints, 3 joint constraints, and 1 global constraint;
 \item For the \BFM: 5 individual constraints, 3 joint constraints, and 1 global constraint.
\end{itemize}
Each constraint was tested by computing the corresponding $p$ as detailed in Section~\ref{ss:meth:inf}.

\subsection{Evaluation} \label{ss:simu:eval}

For each model used to generate the data (\TPM\ and \BFM) and each model providing the constraints tested (again, \TPM\ and \BFM), we computed two summaries:
\begin{itemize}
 \item the 5\% percentile for $p$, i.e., the value $p_{5\%}$ for which we have $p < p_{5\%}$ for 5\% of the 1\,000 samples (i.e., for 50 values);
 \item the fraction $f_{0.05}$ of tests declared significant at a significance level of $\alpha = 0.05$.
\end{itemize}
When analyzing data generated according to a structural model with the constraints associated with the same model (i.e., testing constraints from the \TPM\ using data generated according to the \TPM, or testing constraints from the \BFM\ using data generated according to the \BFM), we expect to have $p_{5\%} \approx 0.05$ and $f_{0.05} \approx 0.05$.

\subsection{Results} \label{ss:simu:res}

Results of simulation study can be found in Table~\ref{tab:simu:res}. When testing the constraints of the \TPM\ on data generated using the \TPM, the values of $p_{5\%}$ and $f_{0.05}$ were both relatively close to the expected value of 0.05 for the individual constraints (range for $p_{5\%}$: 0.029--0.052; range for $f_{0.05}$:  0.029--0.060) and the joint constraints (range: 0.044--0.060), but quite different for the global constraint ($p_{5\%} = 0.252$, $f_{0.05} = 0.004$). When testing the constraints associated with the \TPM\ with data generated using the \BFM, values where much lower for $p_{5\%}$ (range: $< 0.001$ to 0.038) and much larger for $f_{0.05}$ (range 0.072--0.941) for all types of constraints.
\par
By contrast, when testing the constraints associated with the \BFM, we found limited difference in results between data generated using the \TPM\ (range for $p_{5\%}$: 0.0163--0.116; range for $f_{0.05}$:  0.010--0.117, all constraint types) and data generated using the \BFM\ (range for $p_{5\%}$: 0.028--0.148; range for $f_{0.05}$:  0.005--0.072, all constraint types).

\begin{sidewaystable}[!htbp]
 \centering
 \caption{\textbf{Simulation study.} Results of analysis: $p_{5\%}$, the 5\% percentile interval for $p$, as well as $f_{0.05}$, the fraction of samples that were declared significant for a significance level of $\alpha = 0.05$.}\label{tab:simu:res}
 \begin{tabular}{c|l|l|c|c|c|c}
   \multirow{3}{*}{Tested model} & \multirow{3}{*}{Connection} & \multirow{3}{*}{Constraints} & \multicolumn{4}{c}{Generative model} \\
   \hhline{~~~----}
   & & & \multicolumn{2}{c|}{TP} & \multicolumn{2}{c}{BF} \\
   \hhline{~~~----}
   & & & $p_{5\%}$ & $f_{0.05}$ & $p_{5\%}$ & $f_{0.05}$ \\
   \hline
   TP & VEC--SMA & $C_1 \quad \corr [ \vec, \sma | \pfc, \ifg ] = 0$ & 0.052 & 0.049 & 0.002 & 0.231 \\
      &          & $C_2 \quad \corr [ \vec, \sma | \pfc, \ifg, \ipl ] = 0$ & 0.041 & 0.063 & 0.027 & 0.072 \\
   \hhline{~~-----}
      &          & $J_1 \quad$ joint constraint & 0.044 & 0.053 & $< 0.001$ & 0.550 \\
   \hhline{~------}
      & VEC--IFG & none & --- & --- & --- & --- \\
   \hhline{~------}
      & PFC--IFG & $C_3 \quad \corr [ \pfc, \ifg | \vec, \sma ] = 0$ & 0.038 & 0.059 & $< 0.001$ & 0.926 \\
      &          & $C_4 \quad \corr [ \pfc, \ifg | \vec, \sma, \ipl ] = 0$ & 0.042 & 0.057 & $< 0.001$ & 0.859 \\
   \hhline{~~-----}
      &          & $J_2 \quad$ joint constraint & 0.060 & 0.043 & $< 0.001$ & 0.879 \\
   \hhline{~------}
      & PFC--IPL & $C_5 \quad \corr [ \pfc, \ipl | \vec, \ifg ] = 0$ & 0.034 & 0.067 & 0.002 & 0.259 \\
      &          & $C_6 \quad \corr [ \pfc, \ipl | \vec, \sma ] = 0$ & 0.044 & 0.061 & 0.001 & 0.340 \\
      &          & $C_7 \quad \corr [ \pfc, \ipl | \vec, \sma, \ifg] = 0$ & 0.039 & 0.061 & 0.038 & 0.063 \\
   \hhline{~~-----}
      &         & $J_3 \quad$ joint constraint & 0.058 & 0.040 & $< 0.001$ & 0.941 \\
   \hhline{~------}
      & SMA--IPL & $C_8 \quad \corr [ \sma, \ipl | \pfc, \ifg ] = 0$ & 0.037 & 0.067 & $< 0.001$ & 0.743 \\
      &          & $C_9 \quad \corr [ \sma, \ipl | \vec, \ifg ] = 0$ & 0.029 & 0.075 & $< 0.001$ & 0.749 \\
      &          & $C_{10} \quad \corr [ \sma, \ipl | \vec, \pfc, \ifg ] = 0$ & 0.029 & 0.072 & $< 0.001$ & 0.638 \\
   \hhline{~~-----}
      &          & $J_4 \quad$ joint constraint & 0.048 & 0.054 & $< 0.001$ & 0.712 \\
   \hhline{~------}
      &          & $G \quad$ global constraint & 0.252 & 0.004 & 0.004 & 0.534 \\
   \hline
   \hline
   BF & VEC--SMA & $C_1' \quad \corr [ \vec, \sma | \pfc, \ipl ] = 0$ & 0.040 & 0.061 & 0.033 & 0.071 \\
      &          & $C_2' \quad \corr [ \vec, \sma | \pfc, \ifg, \ipl ] = 0$ & 0.041 & 0.062 & 0.028 & 0.072 \\
   \hhline{~~-----}
      &          & $J_1' \quad$ joint constraint & 0.060 & 0.040 & 0.082 & 0.026 \\
   \hhline{~------}
      & VEC--IFG & $C_3' \quad \corr [ \vec, \ifg | \pfc, \ipl ] = 0$ & 0.021 & 0.117 & 0.050 & 0.050 \\
      &          & $C_4' \quad \corr [ \vec, \ifg | \pfc, \sma, \ipl ] = 0$ & 0.016 & 0.113 & 0.051 & 0.049 \\
   \hhline{~~-----}
      &          & $J_2' \quad$ joint constraint & 0.041 & 0.059 & 0.106 & 0.018 \\
   \hhline{~------}
      & PFC--IPL & $C_5' \quad \corr [ \pfc, \ipl | \vec, \sma, \ifg] = 0$ & 0.040 & 0.061 & 0.038 & 0.063 \\
   \hhline{~~-----}
      &          & $J_3' \quad$ joint constraint & 0.040 & 0.061 & 0.038 & 0.063 \\
   \hhline{~------}
      & SMA--IFG & none & --- & --- & --- & --- \\
   \hhline{~------}
      &          & $G' \quad$ global constraint & 0.116 & 0.010 & 0.148 & 0.005
 \end{tabular}
\end{sidewaystable}

\section{Experimental data} \label{s:reel}

We finally used the experimental data provided in \gcitet{Bullmore-2000} to reanalyze their study and provide new insight into the structural modeling.

\subsection{Data}

Each of the 5 regions mentioned in Section~\ref{ss:meth:ex} was associated with a time course of length $T=96$ time samples. The sample correlation matrix between the time series was given in \gcitet{Bullmore-2000}  and is reported in Table~\ref{tab:reel:corr}.
\par

\begin{table}[!hbtp]
  \centering
  \caption{\textbf{Experimental data.} Sample correlation matrix of the data set examined in \gcitet{Bullmore-2000} (lower triangular matrix) and corresponding sample partial correlation matrix (upper triangular matrix, italics).} \label{tab:reel:corr}
  \begin{tabular}{c|ccccc}
        & VEC   & PFC   & SMA   & IFG   & IPL \\
    \hline
    VEC &       &  \textit{0.305} & \textit{0.023} & \textit{0.089} & \textit{0.495} \\
    PFC & 0.661 &                 & \textit{0.420} & \textit{0.164} & \textit{0.132} \\
    SMA & 0.525 & 0.660           &                & \textit{0.091} & \textit{0.170} \\
    IFG & 0.486 & 0.507           & 0.437          &                & \textit{0.188} \\
    IPL & 0.731 & 0.630           & 0.558          & 0.517          &
  \end{tabular}
\end{table}

\subsection{Analysis}

We applied our procedure to the real data to infer in what measure they support the existence of the \TPM\ and/or the \BFM. In particular, while the \TPM\ and the \BFM\ differed both structurally (different set of arrows) and numerically (different path coefficients for arrows that are common to both models), \gcitet{Bullmore-2000} concluded that the data did not contain enough evidence to enable one to discard the theoretically preferred model as being significantly different from the best fit model. We wanted to qualify this statement.

\subsection{Results}

\begin{figure}[!htbp]
  \centering
  \begin{tabular}{cc}
    Conditional correlation & Divergence \\
    \includegraphics[width=0.4\linewidth]{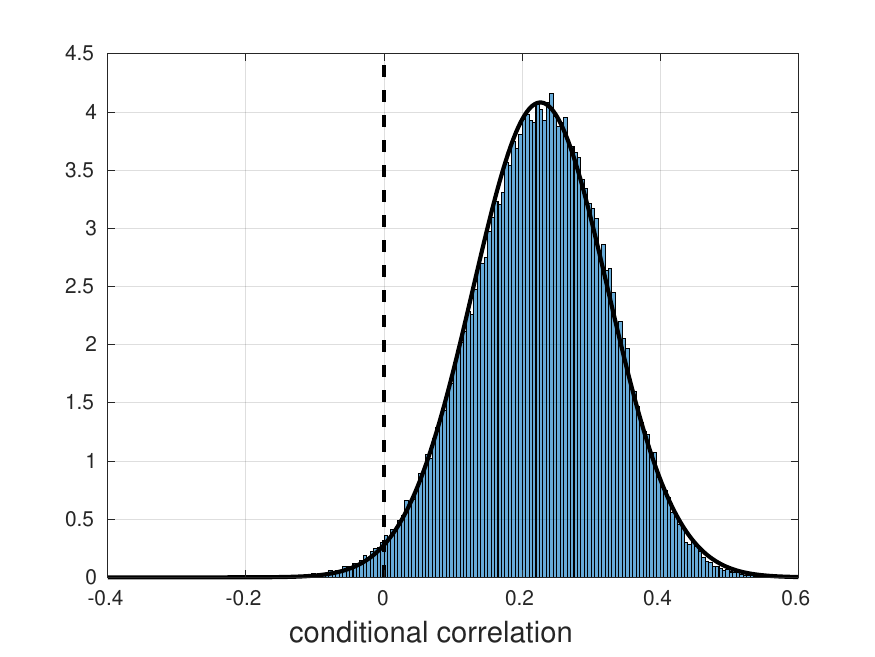}
    & \includegraphics[width=0.4\linewidth]{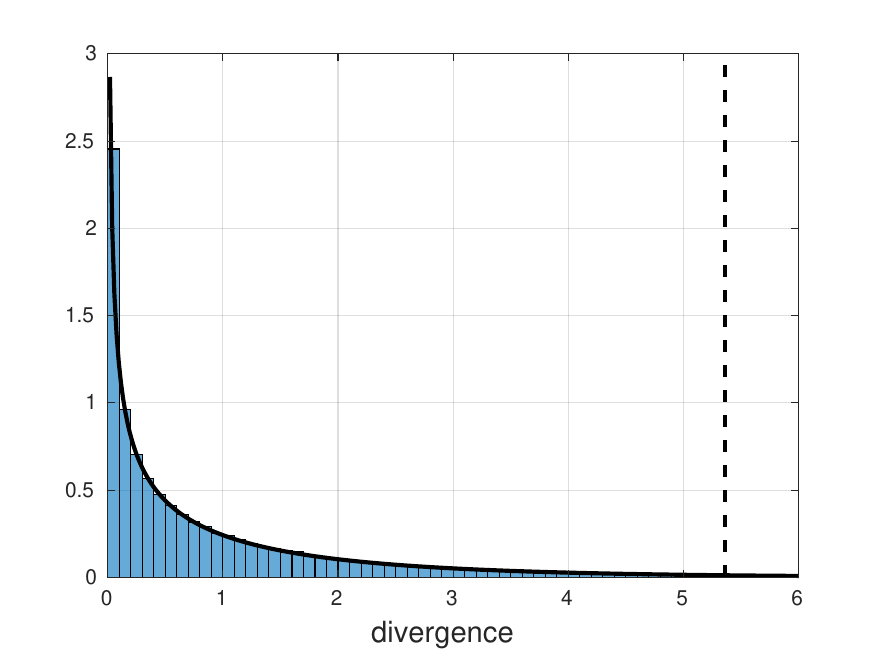} \\
    \includegraphics[width=0.4\linewidth]{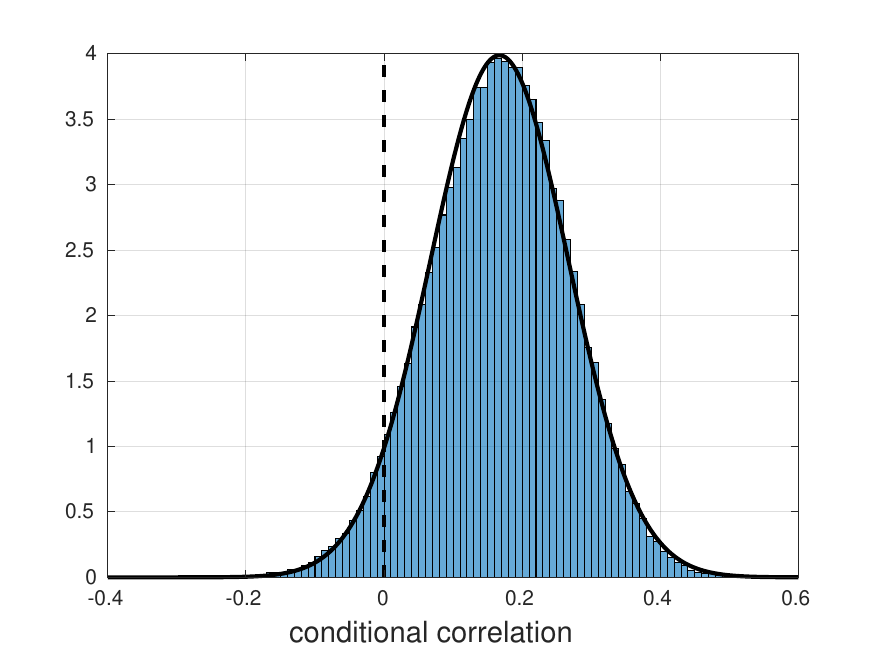}
    & \includegraphics[width=0.4\linewidth]{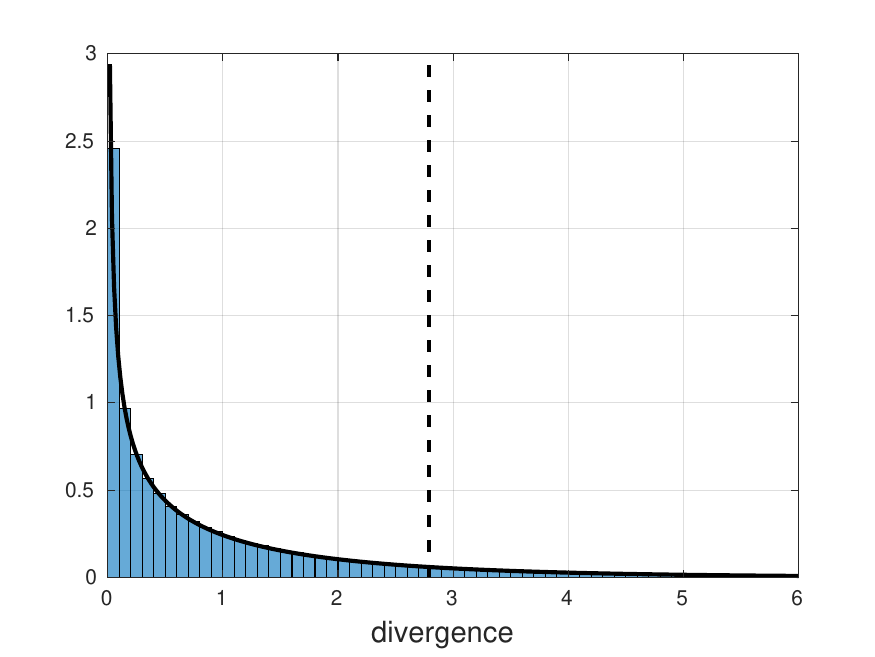} \\
    \includegraphics[width=0.4\linewidth]{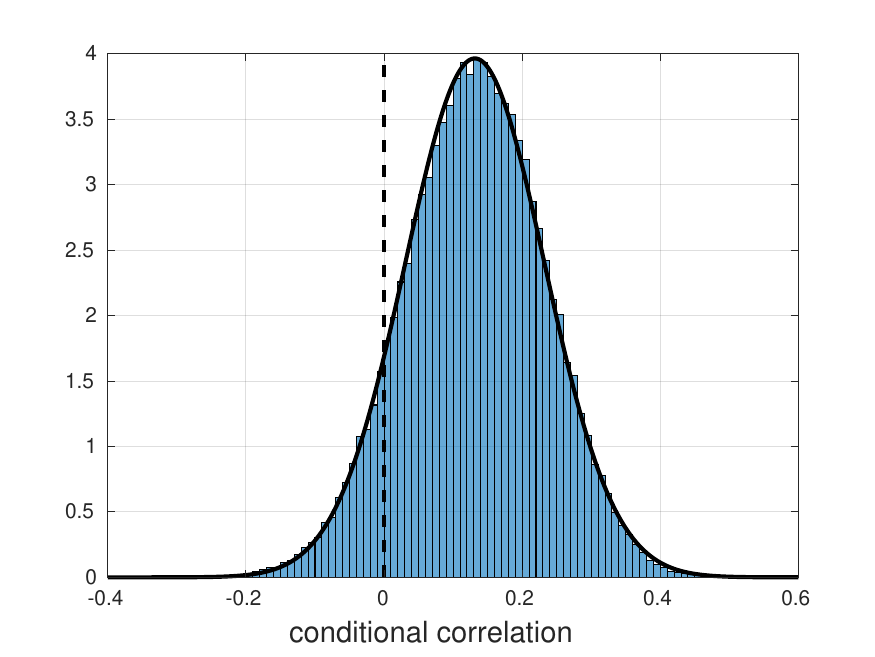}
    & \includegraphics[width=0.4\linewidth]{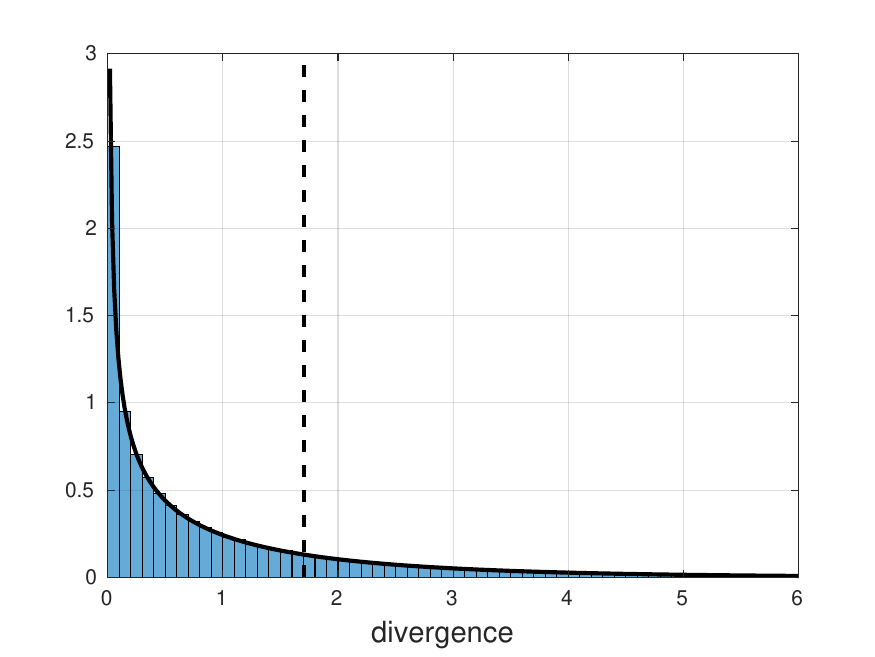}
  \end{tabular}
  \caption{\textbf{Experimental data.} Results from inference corresponding to the three individual constraints relative to the missing link between PFC and IPL in the \TPM: $C_5$ (top), $C_6$ (middle) and $C_7$ (bottom). Left: histograms of the samples approximating the posterior distributions of the conditional correlation coefficients, and corresponding normal approximations. Right: histograms of the samples approximating the posterior distributions of the deviances obtained from Equation~\eqref{eq:def:dev}, and corresponding chi-square approximations.} \label{fig:reel:res:ex1}
\end{figure}

\begin{figure}[!htbp]
  \centering
  \begin{tabular}{cc}
    \includegraphics[width=0.5\linewidth]{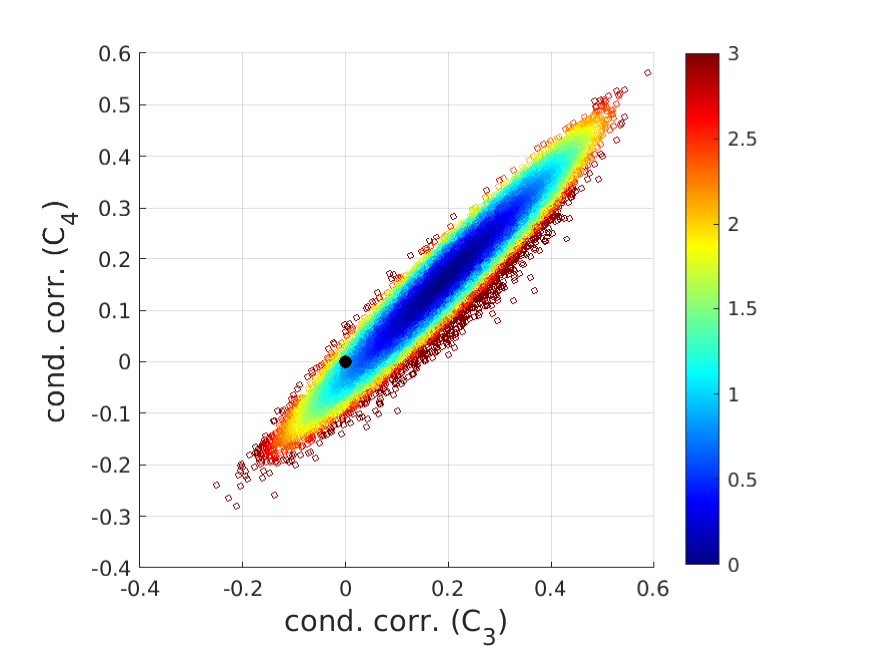}
    & \includegraphics[width=0.5\linewidth]{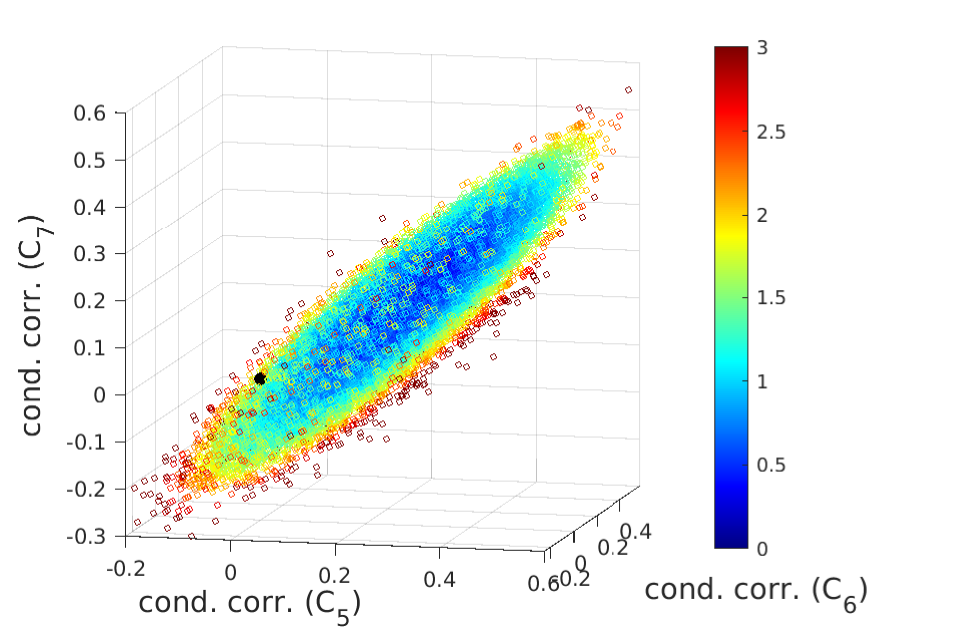}
  \end{tabular}
  \caption{\textbf{Experimental data.} Results from inference corresponding to the joint constraints relative to the missing link between PFC and IFG (left) as well as between PFC and IPL (right) in the \TPM. Scatterplot of the conditional correlation coefficients associated with $J_2$ and $J_3$. The color codes $- \log_{10} d ( \tilde{\vect{\rho}} )$, thresholded at 3. The black dot stands for $\tilde{\vect{\rho}} = \vect{0}$.} \label{fig:reel:res:ex2}
\end{figure}

See Figures \ref{fig:reel:res:ex1} and \ref{fig:reel:res:ex2} for an example of output from the approximate sampling scheme. The significance of the different constraints are reported in Table~\ref{tab:reel:res:res1} and summarized in graphical form in Figure~\ref{fig:reel:res:res2}. While the global set of constraints could not be rejected for the \TPM\ ($p = 0.171$), some specific constraints (such as $C_5$, $C_8$ and $C_9$; also $C_3$ close to significance), or sets of constraints (such as $J_3$, corresponding to a lack of connection between PFC and IPL  and $J_4$, corresponding to a lack of connection between SMA and IPL), were found to be significantly different from 0. By contrast, in the \BFM, no individual, joint or global constraint could be rejected.

\begin{table}[!htbp]
 \centering
 \caption{\textbf{Experimental data.} Result of inference: significance values for the different constraints entailed by the \TPM\ and the \BFM. Values lower than a threshold of $\alpha = 0.05$ are emphasized in bold, while values larger but close to $\alpha$ are in italics.} \label{tab:reel:res:res1}
 \begin{tabular}{c|l|l|c}
   Tested model & Connection & Constraints & $p$ \\
   \hline
   TP & VEC--SMA & $C_1 \quad \corr [ \vec, \sma | \pfc, \ifg ] = 0$ & 0.220 \\
      &          & $C_2 \quad \corr [ \vec, \sma | \pfc, \ifg, \ipl ] = 0$ & 0.823 \\
   \hhline{~~--}
      &          & $J_1 \quad$ joint constraint & 0.136 \\
   \hhline{~---}
      & VEC--IFG & none & --- \\
   \hhline{~---}
      & PFC--IFG & $C_3 \quad \corr [ \pfc, \ifg | \vec, \sma ] = 0$ & \textit{0.052} \\
      &          & $C_4 \quad \corr [ \pfc, \ifg | \vec, \sma, \ipl ] = 0$ & 0.105 \\
   \hhline{~~--}
      &          & $J_2 \quad$ joint constraint & 0.098 \\
   \hhline{~---}
      & PFC--IPL & $C_5 \quad \corr [ \pfc, \ipl | \vec, \ifg ] = 0$ & \textbf{0.020} \\
      &          & $C_6 \quad \corr [ \pfc, \ipl | \vec, \sma ] = 0$ & 0.094 \\
      &          & $C_7 \quad \corr [ \pfc, \ipl | \vec, \sma, \ifg] = 0$ & 0.192 \\
   \hhline{~~--}
      &         & $J_3 \quad$ joint constraint & \textbf{0.017} \\
   \hhline{~---}
      & SMA--IPL & $C_8 \quad \corr [ \sma, \ipl | \pfc, \ifg ] = 0$ & \textbf{0.034} \\
      &          & $C_9 \quad \corr [ \sma, \ipl | \vec, \ifg ] = 0$ & \textbf{0.009} \\
      &          & $C_{10} \quad \corr [ \sma, \ipl | \vec, \pfc, \ifg ] = 0$ & 0.089 \\
   \hhline{~~--}
      &          & $J_4 \quad$ joint constraint & \textbf{0.014} \\
   \hhline{~---}
      &          & $G \quad$ global constraint & 0.171 \\
   \hline
   \hline
   BF & VEC--SMA & $C_1' \quad \corr [ \vec, \sma | \pfc, \ipl ] = 0$ & 0.765 \\
      &          & $C_2' \quad \corr [ \vec, \sma | \pfc, \ifg, \ipl ] = 0$ & 0.830 \\
   \hhline{~~--}
      &          & $J_1' \quad$ joint constraint & 0.828 \\
   \hhline{~---}
      & VEC--IFG & $C_3' \quad \corr [ \vec, \ifg | \pfc, \ipl ] = 0$ & 0.380 \\
      &          & $C_4' \quad \corr [ \vec, \ifg | \pfc, \sma, \ipl ] = 0$ & 0.340 \\
   \hhline{~~--}
      &          & $J_2' \quad$ joint constraint & 0.588 \\
   \hhline{~---}
      & PFC--IPL & $C_5' \quad \corr [ \pfc, \ipl | \vec, \sma, \ifg] = 0$ & 0.188 \\
   \hhline{~~--}
      &          & $J_3' \quad$ joint constraint & 0.188 \\
   \hhline{~---}
      & SMA--IFG & none & --- \\
   \hhline{~---}
      &          & $G' \quad$ global constraint & 0.690
 \end{tabular}
\end{table}

\begin{figure}[!htb]
  \centering
  \begin{tabular}{cc}
   Theoretically preferred model (TP) & Best fit model (BF) \\
   \begin{tikzpicture}[scale=0.8]
    \node[region] (pfc) at (0,0) {PFC};
    \node[region] (sma) at (2,2) {SMA};
    \node[region] (ifg) at (3,-0.7) {IFG};
    \node[region] (ipl) at (5.5,0.7) {IPL};
    \node[region] (vec) at (5.5,-2) {VEC};
    \draw[lien] (pfc) -- (sma);
    \draw[lien] (sma) -- (ifg);
    \draw[lien] (ifg) -- (ipl);
    \draw[lien] (ipl) to[bend right] (vec);
    \draw[lien] (vec) to[bend right] (ipl);
    \draw[lien] (vec.south west) to[bend left=45] (pfc);
    \draw[line width= 0.7mm,blue!50,loosely dotted] (vec) to[bend left=5] (sma);
    \draw[line width= 0.7mm,orange!50,,dash pattern=on 3mm off 3mm] (pfc) to (ifg);
    \draw[line width= 0.7mm,red] (pfc) to (ipl);
    \draw[line width= 0.7mm,red] (sma) to (ipl);
   \end{tikzpicture}
   &
   \begin{tikzpicture}[scale=0.8]
    \node[region] (pfc) at (0,0) {PFC};
    \node[region] (sma) at (2,2) {SMA};
    \node[region] (ifg) at (3,-0.7) {IFG};
    \node[region] (ipl) at (5.5,0.7) {IPL};
    \node[region] (vec) at (5.5,-2) {VEC};
    \draw[lien] (pfc) -- (sma);
    \draw[lien] (pfc) -- (ifg);
    \draw[lien] (sma) -- (ipl);
    \draw[lien] (ifg) -- (ipl);
    \draw[lien] (ipl) --  (vec);
    \draw[lien] (vec.south west) to[bend left=45] (pfc);
    \draw[line width= 0.7mm,blue!50,loosely dotted] (vec) to (sma);
    \draw[line width= 0.7mm,blue!50,loosely dotted] (vec) to (ifg);
    \draw[line width= 0.7mm,blue!50,loosely dotted] (pfc) to (ipl);
   \end{tikzpicture} \\
   Partial correlation \\
   \begin{tikzpicture}[scale=0.8]
    \node[region] (pfc) at (0,0) {PFC};
    \node[region] (sma) at (2,2) {SMA};
    \node[region] (ifg) at (3,-0.7) {IFG};
    \node[region] (ipl) at (5.5,0.7) {IPL};
    \node[region] (vec) at (5.5,-2) {VEC};
    \draw[line width= 0.7mm,red] (pfc) -- (sma);
    \draw[line width= 0.7mm,red] (ipl) --  (vec);
    \draw[line width= 0.7mm,red] (vec.south west) to[bend left=45] (pfc);
    \draw[line width= 0.7mm,orange,dash pattern=on 3mm off 3mm] (sma) -- (ipl);
    \draw[line width= 0.7mm,orange,dash pattern=on 3mm off 3mm] (ifg) -- (ipl);
    \draw[line width= 0.7mm,orange,dash pattern=on 3mm off 3mm] (pfc) -- (ifg);
    \draw[line width= 0.7mm,blue!50,loosely dotted] (vec) to (sma);
    \draw[line width= 0.7mm,blue!50,loosely dotted] (vec) to (ifg);
    \draw[line width= 0.7mm,blue!50,loosely dotted] (pfc) to (ipl);
    \draw[line width= 0.7mm,blue!50,loosely dotted] (sma) -- (ifg);
   \end{tikzpicture}
  \end{tabular}
  \caption{\textbf{Experimental data.} Graphical representation of the values of  $p$ associated with missing links (joint constraints) for the \TPM\ (TP, top left) and \BFM\ (BF, top right). Undirected links correspond to missing arrows and values of $p$ associated with the corresponding joint constraints. Bottom: values of $p$ for  partial correlations calculated using deviance. Blue dotted lines: $p \geq 0.1$; orange dashed lines: $0.05 \leq p < 0.11$; solid red line: $p < 0.05$.} \label{fig:reel:res:res2}
\end{figure}

\section{Discussion} \label{s:disc}

In the present paper, we proposed a novel method to test the validity of a given model of structural equation. Given a structural model in the form of a directed (cyclic or acylic) graph, it extracts the set of all constraints of conditional independence induced by the absence of links between regions in the model and tests for their validity in a Bayesian framework, either individually (constraint by constraint), jointly (by gathering all constraints associated with a given missing link), or globally (all constraints associated with a structural model). We illustrated the approach on a dataset and two structural models. With a simulation study, we showed the power and limits of the method. Finally, we applied the method to the real data.
\par
This approach is unique in that it avoids several issues that are typical of usual SEM inference methods. First, it does not mislead the user into making incorrect conclusions regarding the causal pattern of the model, as observationally equivalent models will lead to the same conclusion. Second, it makes it possible to test constraints at the level desired by the user: either at the scale of a single constraint, a set of constraints corresponding to a missing link, and all constraints specific of the structural model.
\par
\begin{figure}[!htbp]
  \centering
  \begin{tabular}{cc}
    Theoretically preferred (TP) & Best fit (BF) \\
    \includegraphics[width=0.5\linewidth]{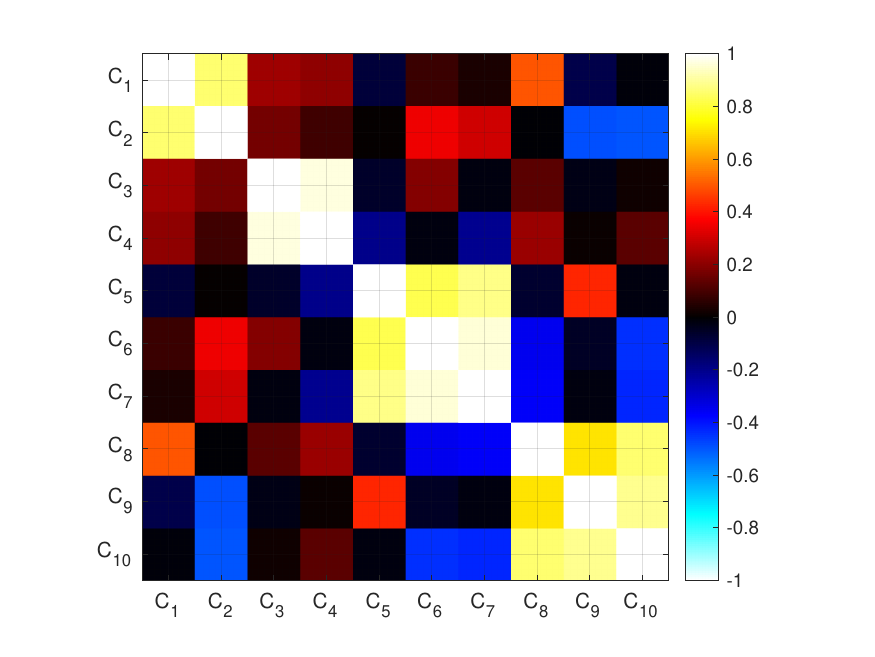}
    &
    \includegraphics[width=0.5\linewidth]{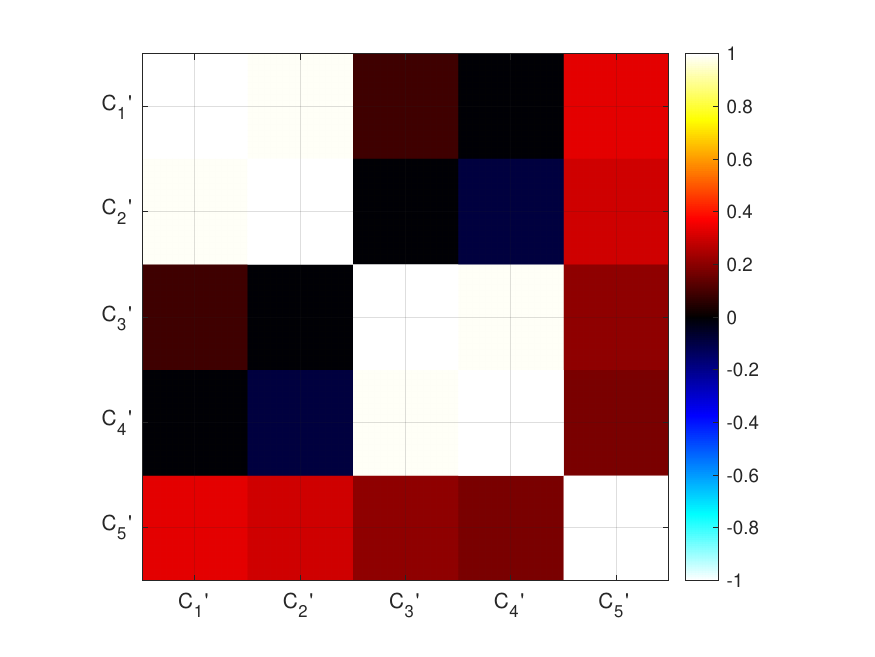}
  \end{tabular}
  \caption{\textbf{Experimental data.} Value of correlation between conditional correlation coefficients approximated with the sampling scheme.} \label{fig:reel:res:correl}
\end{figure}

A key point in the procedure is the determination of the constraints of conditional independence entailed by the structural model \gcite{Acid-2013}. Efficient methods have been proposed to extract such constraints in a principled fashion. For acyclic graphs, a method was proposed by \gcitet{Pearl-1988}. Another, more general approach was introduced by \gcitet{Shipley-2000}. Both approaches rely on the fact that the set of all constraints entailed by a structural model are not necessarily independent from one another, as some statements can be predicted from others. For instance, we could expect constraints involving the same pair of variables $( i, j )$ to be highly correlated. The two methods mentioned above were developed to avoid this redundancy and to extract a (not necessarily unique) smallest subset, called basis, which still generates all existing relationships. We here departed from this approach and, instead, used brute force to extract and deal with \emph{all} relationships of conditional independence. We reasoned that the redundancy contained in the set of all constraints could be beneficial to the inference process. Indeed, discarding some constraints is tantamount to discarding information about the model, which is not desired. Also, when dealing with real data, we can expect some constraints to be better respected than others in the data; which constraints are best to investigate a model is therefore the result of an interaction between model and data---i.e., it is itself an inference process, which we did not take the risk of separating from the main task. The application we introduced confirmed the strong redundancy between constraints related to the same missing link (e.g., for the \TPM: $C_3$ and $C_4$, corresponding to the missing connection between PFC and IFG; $C_5$, $C_6$, and $C_7$, corresponding to the missing connection between PFC and IPL), as observed by visual inspection of either the scatterplots (see Figure~\ref{fig:reel:res:ex2}) or the correlation matrices (see Figure~\ref{fig:reel:res:correl}). The application also brought evidence in favor of keeping all constraints, as various relationships corresponding to the same missing link were found to have quite different significance levels. For instance, in the \TPM, $C_5$, $C_6$ and $C_7$, related to the same missing connection between PFC and IPL, had $p$-values ranging from 0.020 to 0.192.
\par
This exhaustive way of extracting constraints may limit the generalization of the method to larger systems, as its scalability with respect to the number of regions remains to be investigated. In particular, we are not sure whether there exists an efficient algorithm that extracts all relationships of conditional independence from a given structural model.
\par
Note that, in the process of extracting constraints of conditional independence, a missing link does not necessarily translate into a constraint. This is in particular true in cyclic graphs, i.e., in the presence of loops or feedbacks. For instance, the absence of a link between IFG and VE in the \TPM\ was not associated with a constraint. Similarly, in the \BFM, the missing link between SMA and IFG was not associated with a constraint as their unique parent PFC was also a descendant.

\par
To assess the validity of our approach, we used a simulation study where we generated data according to the two models under investigation (\TPM\ and \BFM). Interestingly, this showed evidence of a differential behavior depending on the model being tested. Our method was able to discriminate synthetic data generated according to the \TPM\ from synthetic data generated from the \BFM\ in the case of constraints associated with the \TPM. By contrast, it was not able to do so in the case of constraints associated with the \BFM.
\par
The approach introduced in the present manuscript strongly relies on conditional correlation. Conditional correlation was introduced in fMRI functional connectivity analysis as a data-driven way to extract information from data that is closer to effective connectivity than pairwise correlation (\gcitep{Marrelec-2005b}a, \gcitep{Marrelec-2005c}b). A particular emphasis was put on partial correlation, a particular type of conditional correlation where the conditioning set is the rest of the variables \gcite{Marrelec-2006c, Marrelec-2007, Marrelec-2009, Marrelec-2009b, Smith_SM-2011}. By contrast, the present method is clearly model-based, as it requires the knowledge of a potential model to start with.
\par
For instance, consider PFC and IPL. In the \TPM, the absence of a link between both regions was deemed to be implausible in the face of the data ($p = 0.017$), while it was found to agree with the data based on the \BFM\ ($p = 0.188$).
\par
The validity of using SEM in fMRI as well as its strengths and limits, and even the very possibility to access information relative to effective connectivity, are topics that are still discussed \gcite{Goncalves-2001, Goncalves-2003, Ramsey_JD-2010, Friston-2011b}. Still, SEM is a classical way to deal with effective connectivity, and we reasoned that improving this method such that it avoids the pitfall of observationally equivalent models and provides for a finer view into local relationships could only be beneficial to the community of SEM users. We also hope that our contribution will broaden the usability of SEM and help the neuroscientists improve the causal information that they can extract from data.

\bibliographystyle{chicago} 
\bibliography{nonabrev,anglais,mabiblio}

\begin{thebibliography}{}

\bibitem[\protect\citeauthoryear{Acid and {de Campos}}{Acid and {de
  Campos}}{2013}]{Acid-2013}
Acid, S. and L.~M. {de Campos} (2013).
\newblock An algorithm for finding minimum d-separating sets in belief
  networks.
\newblock arXiv:1302.3549 [cs.AI].

\bibitem[\protect\citeauthoryear{Anderson}{Anderson}{1958}]{Anderson_TW-1958}
Anderson, T.~W. (1958).
\newblock {\em An Introduction to Multivariate Statistical Analysis}.
\newblock Wiley Publications in Statistics. John Wiley and Sons, New York.

\bibitem[\protect\citeauthoryear{Barrett and Barnett}{Barrett and
  Barnett}{2013}]{Barrett-2013}
Barrett, A.~B. and L.~Barnett (2013).
\newblock Granger causality is designed to measure effect, not mechanism.
\newblock {\em Frontiers in Neuroinformatics\/}~{\em 7}, 6.

\bibitem[\protect\citeauthoryear{Bielczyk, Uithol, {van Mourik}, Anderson,
  Glennon, and Buitelaar}{Bielczyk et~al.}{2019}]{Bielczyk-2019}
Bielczyk, N.~Z., S.~Uithol, T.~{van Mourik}, P.~Anderson, J.~C. Glennon, and
  J.~K. Buitelaar (2019).
\newblock Disentangling causal webs in the brain using functional magnetic
  resonance imaging: a review of current approaches.
\newblock {\em Network Neuroscience\/}~{\em 3\/}(2), 237--273.

\bibitem[\protect\citeauthoryear{Bollen}{Bollen}{1989}]{Bollen-1989}
Bollen, K.~A. (1989).
\newblock {\em Structural Equation with Latent Variables}.
\newblock Wiley-interscience.

\bibitem[\protect\citeauthoryear{B{\"u}chel and Friston}{B{\"u}chel and
  Friston}{1997}]{Buchel-1997}
B{\"u}chel, C. and K.~J. Friston (1997).
\newblock Modulation of connectivity in visual pathways by attention: cortical
  interactions evaluated with structural equation modelling and f{MRI}.
\newblock {\em Cerebral Cortex\/}~{\em 7}, 768--778.

\bibitem[\protect\citeauthoryear{Bullmore, Horwitz, Honey, Brammer, Williams,
  and Sharma}{Bullmore et~al.}{2000}]{Bullmore-2000}
Bullmore, E., B.~Horwitz, G.~Honey, M.~Brammer, S.~Williams, and T.~Sharma
  (2000).
\newblock How good is good enough in path analysis of f{MRI} data?
\newblock {\em NeuroImage\/}~{\em 11}, 289--301.

\bibitem[\protect\citeauthoryear{Civelek}{Civelek}{2018}]{Civelek-2018}
Civelek, M.~E. (2018).
\newblock {\em Essentials of Structural Equation Modeling}.
\newblock Zea E-Books, Lincoln, NE.

\bibitem[\protect\citeauthoryear{Cudeck, Klebe, and Henly}{Cudeck
  et~al.}{1993}]{Cudeck-1993}
Cudeck, R., K.~J. Klebe, and S.~J. Henly (1993).
\newblock A simple {G}auss-{N}ewton procedure for covariance structure analysis
  with high-level computer languages.
\newblock {\em Psychometrika\/}~{\em 58}, 211--232.

\bibitem[\protect\citeauthoryear{Daunizeau, David, and Stephan}{Daunizeau
  et~al.}{2011}]{Daunizeau-2011}
Daunizeau, J., O.~David, and K.~E. Stephan (2011).
\newblock Dynamic causal modelling: a critical review of the biophysical and
  statistical foundations.
\newblock {\em NeuroImage\/}~{\em 58\/}(2), 312--322.

\bibitem[\protect\citeauthoryear{Davey, Grayden, Gavrilescu, Egan, and
  Johnston}{Davey et~al.}{2013}]{Davey-2013}
Davey, C.~E., D.~B. Grayden, M.~Gavrilescu, G.~F. Egan, and L.~A. Johnston
  (2013).
\newblock The equivalence of linear gaussian connectivity techniques.
\newblock {\em Human Brain Mapping\/}~{\em 34\/}(9), 1999--2014.

\bibitem[\protect\citeauthoryear{Davvetas, Diamantopoulos, Zaefarian, and
  Sichtmann}{Davvetas et~al.}{2020}]{Davvetas-2020}
Davvetas, V., A.~Diamantopoulos, G.~Zaefarian, and C.~Sichtmann (2020).
\newblock Ten basic questions about structural equations modeling you should
  know the answers to -- {B}ut perhaps you don't.
\newblock {\em Industrial Marketing Management\/}~{\em 90}, 252--263.

\bibitem[\protect\citeauthoryear{Friston, Preller, Mathys, Cagnan, Heinzle,
  Razi, and Zeidman}{Friston et~al.}{2017}]{Friston-2017}
Friston, K., K.~H. Preller, C.~Mathys, H.~Cagnan, J.~Heinzle, A.~Razi, and
  P.~Zeidman (2017).
\newblock Dynamic causal modelling revisited.
\newblock {\em NeuroImage\/}~{\em 199}, 730--744.

\bibitem[\protect\citeauthoryear{Friston}{Friston}{1994}]{Friston-1994b}
Friston, K.~J. (1994).
\newblock Functional and effective connectivity in neuroimaging: a synthesis.
\newblock {\em Human Brain Mapping\/}~{\em 2\/}(1--2), 56--78.

\bibitem[\protect\citeauthoryear{Friston}{Friston}{2011}]{Friston-2011b}
Friston, K.~J. (2011).
\newblock Functional and effective connectivity: a review.
\newblock {\em Brain Connectivity\/}~{\em 1\/}(1), 13--36.

\bibitem[\protect\citeauthoryear{Friston, Harrison, and Penny}{Friston
  et~al.}{2003}]{Friston-2003b}
Friston, K.~J., L.~Harrison, and W.~Penny (2003).
\newblock Dynamic causal modelling.
\newblock {\em NeuroImage\/}~{\em 19\/}(4), 1273--1302.

\bibitem[\protect\citeauthoryear{Gelman, Carlin, Stern, and Rubin}{Gelman
  et~al.}{1998}]{Gelman-1998}
Gelman, A., J.~B. Carlin, H.~S. Stern, and D.~B. Rubin (1998).
\newblock {\em Bayesian Data Analysis}.
\newblock Texts in Statistical Science. Chapman \& Hall, London.

\bibitem[\protect\citeauthoryear{Goebel, Roebroeck, Kim, and Formisano}{Goebel
  et~al.}{2003}]{Goebel-2003}
Goebel, R., A.~Roebroeck, D.-S. Kim, and E.~Formisano (2003).
\newblock Investigating directed cortical interactions in time-resolved f{MRI}
  data using vector autoregressive modeling and {G}ranger causality mapping.
\newblock {\em Magnetic Resonance Imaging\/}~{\em 21}, 1251--1261.

\bibitem[\protect\citeauthoryear{Gon{\c c}alves, Hall, Johnsrude, and
  Haggard}{Gon{\c c}alves et~al.}{2001}]{Goncalves-2001}
Gon{\c c}alves, M.~S., D.~A. Hall, I.~S. Johnsrude, and M.~P. Haggard (2001).
\newblock Can meaningful effective connectivities be obtained between auditory
  cortical regions?
\newblock {\em NeuroImage\/}~{\em 14}, 1353--1360.

\bibitem[\protect\citeauthoryear{Gon\c{c}alves and Hall}{Gon\c{c}alves and
  Hall}{2003}]{Goncalves-2003}
Gon\c{c}alves, M.~S. and D.~A. Hall (2003).
\newblock Connectivity analysis with structural equation modelling: an example
  of the effects of voxel selection.
\newblock {\em NeuroImage\/}~{\em 20}, 1455--1467.

\bibitem[\protect\citeauthoryear{Gonzalez-Lima and McIntosh}{Gonzalez-Lima and
  McIntosh}{1995}]{Gonzalez-Lima-1995}
Gonzalez-Lima, F. and A.~R. McIntosh (1995).
\newblock Analysis of neural interactions related to associative learning using
  structural equation modeling.
\newblock {\em Mathematics and Computers in Simulation\/}~{\em 40}, 115--140.

\bibitem[\protect\citeauthoryear{Hoyle}{Hoyle}{2012}]{Hoyle-2012}
Hoyle, R.~H. (Ed.) (2012).
\newblock {\em Handbook of Structural Equation Modeling}.
\newblock Guilford Publications.

\bibitem[\protect\citeauthoryear{Hu, Dai, Worrell, Dai, and Liang}{Hu
  et~al.}{2011}]{Hu-2011}
Hu, S., G.~Dai, G.~A. Worrell, Q.~Dai, and H.~Liang (2011).
\newblock Causality analysis of neural connectivity: critical examination of
  existing methods and advances of new methods.
\newblock {\em IEEE Transactions on Neural Networks\/}~{\em 22\/}(6), 829--844.

\bibitem[\protect\citeauthoryear{James, Kelley, Craddock, Holtzheimer, Dunlop,
  Nemeroff, Mayberg, and Hu}{James et~al.}{2009}]{James-2009}
James, G.~A., M.~E. Kelley, R.~C. Craddock, P.~E. Holtzheimer, B.~W. Dunlop,
  C.~B. Nemeroff, H.~S. Mayberg, and X.~P. Hu (2009).
\newblock Exploratory structural equation modeling of resting-state f{MRI}:
  Applicability of group models to individual subjects.
\newblock {\em NeuroImage\/}~{\em 45}, 778--787.

\bibitem[\protect\citeauthoryear{Jovellar and Doudet}{Jovellar and
  Doudet}{2019}]{Jovellar-2019}
Jovellar, D.~B. and D.~J. Doudet (2019).
\newblock f{MRI} in non-human primate: a review on factors that can affect
  interpretation and dynamic causal modeling application.
\newblock {\em Frontiers in Neuroscience\/}~{\em 13}, 973.

\bibitem[\protect\citeauthoryear{Kershaw, Ardekani, and Kanno}{Kershaw
  et~al.}{1999}]{Kershaw-1999}
Kershaw, J., B.~A. Ardekani, and I.~Kanno (1999).
\newblock Application of {B}ayesian inference to f{MRI} data analysis.
\newblock {\em IEEE Transactions on Medical Imaging\/}~{\em 18}, 1138--1153.

\bibitem[\protect\citeauthoryear{Lohmann, Erfurth, Muller, and Turner}{Lohmann
  et~al.}{2012}]{Lohmann-2012}
Lohmann, G., K.~Erfurth, K.~Muller, and R.~Turner (2012).
\newblock Critical comments on dynamic causal modelling.
\newblock {\em NeuroImage\/}~{\em 59}, 2322--2329.

\bibitem[\protect\citeauthoryear{Marrelec and Benali}{Marrelec and
  Benali}{2009}]{Marrelec-2009}
Marrelec, G. and H.~Benali (2009).
\newblock A theoretical investigation of the relationship between structural
  equation modeling and partial correlation in functional {MRI} effective
  connectivity.
\newblock {\em Computational Intelligence and Neuroscience\/}~{\em 2009},
  Article ID 369341, 9 pages.

\bibitem[\protect\citeauthoryear{Marrelec, Benali, Ciuciu,
  P{\'e}l{\'e}grini-Issac, and Poline}{Marrelec et~al.}{2003}]{Marrelec-2003b}
Marrelec, G., H.~Benali, P.~Ciuciu, M.~P{\'e}l{\'e}grini-Issac, and J.-B.
  Poline (2003).
\newblock Robust {B}ayesian estimation of the hemodynamic response function in
  event-related {BOLD} f{MRI} using basic physiological information.
\newblock {\em Human Brain Mapping\/}~{\em 19}, 1--17.

\bibitem[\protect\citeauthoryear{Marrelec, Daunizeau, P{\'e}l{\'e}grini-Issac,
  Doyon, and Benali}{Marrelec et~al.}{2005}]{Marrelec-2005b}
Marrelec, G., J.~Daunizeau, M.~P{\'e}l{\'e}grini-Issac, J.~Doyon, and H.~Benali
  (2005a).
\newblock Conditional correlation as a measure of mediated interactivity in
  f{MRI} and {MEG/EEG}.
\newblock {\em IEEE Transactions on Signal Processing\/}~{\em 53}, 3503--3516.

\bibitem[\protect\citeauthoryear{Marrelec, Doyon, P{\'e}l{\'e}grini-Issac, and
  Benali}{Marrelec et~al.}{2005}]{Marrelec-2005c}
Marrelec, G., J.~Doyon, M.~P{\'e}l{\'e}grini-Issac, and H.~Benali (2005b).
\newblock Heading for data-driven measures of effective connectivity in
  functional {MRI}.
\newblock In {\em Proceedings of the International Joint Conference on Neural
  Networks}, pp.\  1528--1533.

\bibitem[\protect\citeauthoryear{Marrelec, Horwitz, Kim,
  P{\'e}l{\'e}grini-Issac, Benali, and Doyon}{Marrelec
  et~al.}{2007}]{Marrelec-2007}
Marrelec, G., B.~Horwitz, J.~Kim, M.~P{\'e}l{\'e}grini-Issac, H.~Benali, and
  J.~Doyon (2007).
\newblock Using partial correlation to enhance structural equation modeling of
  functional {MRI} data.
\newblock {\em Magnetic Resonance Imaging\/}~{\em 25}, 1181--1189.

\bibitem[\protect\citeauthoryear{Marrelec, Kim, Doyon, and Horwitz}{Marrelec
  et~al.}{2009}]{Marrelec-2009b}
Marrelec, G., J.~Kim, J.~Doyon, and B.~Horwitz (2009).
\newblock Large scale neural model validation of partial correlation analysis
  for effective connectivity investigation in functional {MRI}.
\newblock {\em Human Brain Mapping\/}~{\em 30}, 941--950.

\bibitem[\protect\citeauthoryear{Marrelec, Krainik, Duffau,
  P{\'e}l{\'e}grini-Issac, Leh{\'e}ricy, Doyon, and Benali}{Marrelec
  et~al.}{2006}]{Marrelec-2006c}
Marrelec, G., A.~Krainik, H.~Duffau, M.~P{\'e}l{\'e}grini-Issac,
  S.~Leh{\'e}ricy, J.~Doyon, and H.~Benali (2006).
\newblock Partial correlation for functional brain interactivity investigation
  in functional {MRI}.
\newblock {\em NeuroImage\/}~{\em 32}, 228--237.

\bibitem[\protect\citeauthoryear{Mayekawa}{Mayekawa}{1994}]{Mayekawa-1994}
Mayekawa, S. (1994).
\newblock Equivalent path models in linear structural equation models.
\newblock {\em Behaviormetrika\/}~{\em 21\/}(1), 79--96.

\bibitem[\protect\citeauthoryear{McIntosh and Gonzalez-Lima}{McIntosh and
  Gonzalez-Lima}{1994}]{McIntosh-1994}
McIntosh, A.~R. and F.~Gonzalez-Lima (1994).
\newblock Structural equation modeling and its aplication to network analysis
  of functional brain imaging.
\newblock {\em Human Brain Mapping\/}~{\em 2\/}(1--2), 2--22.

\bibitem[\protect\citeauthoryear{Pearl}{Pearl}{1988}]{Pearl-1988}
Pearl, J. (1988).
\newblock {\em Probabilistic Reasoning in Intelligent Systems: Networks of
  Plausible Inference}.
\newblock Morgan Kaufmann, San Mateo, CA.

\bibitem[\protect\citeauthoryear{Pearl}{Pearl}{2001a}]{Pearl-2001}
Pearl, J. (2001a).
\newblock {\em Causality: Models, Reasoning, and Inference}.
\newblock Cambridge University Press, Cambridge.

\bibitem[\protect\citeauthoryear{Pearl}{Pearl}{2001b}]{Pearl-2001d}
Pearl, J. (2001b).
\newblock Direct and indirect effects.
\newblock arXiv:1301.2300 [cs.AI].

\bibitem[\protect\citeauthoryear{Ramsey, Hanson, Hanson, Halchenko, Poldrack,
  and Glymour}{Ramsey et~al.}{2010}]{Ramsey_JD-2010}
Ramsey, J.~D., S.~J. Hanson, C.~Hanson, Y.~O. Halchenko, R.~A. Poldrack, and
  C.~Glymour (2010).
\newblock Six problems for causal inference from f{MRI}.
\newblock {\em NeuroImage\/}~{\em 49}, 1545--1558.

\bibitem[\protect\citeauthoryear{Roebroeck, Formisano, and Goebel}{Roebroeck
  et~al.}{2005}]{Roebroeck-2005}
Roebroeck, A., E.~Formisano, and R.~Goebel (2005).
\newblock Mapping directed influence over the brain using {G}ranger causality
  and f{MRI}.
\newblock {\em NeuroImage\/}~{\em 25}, 230--242.

\bibitem[\protect\citeauthoryear{Shipley}{Shipley}{2000a}]{Shipley-2000b}
Shipley, B. (2000a).
\newblock {\em Cause and Correlation in Biology --A User's Guide to Path
  Analysis, Structural Equations and Causal Inference}.
\newblock Cambridge University Press.

\bibitem[\protect\citeauthoryear{Shipley}{Shipley}{2000b}]{Shipley-2000}
Shipley, B. (2000b).
\newblock A new inferential test for path models based on directed acyclic
  graphs.
\newblock {\em Structural Equation Modeling\/}~{\em 7}, 206--218.

\bibitem[\protect\citeauthoryear{Smith, Miller, Salimi-Khorshidi, Webster,
  Beckmann, Nichols, Ramsey, and Woolrich}{Smith et~al.}{2011}]{Smith_SM-2011}
Smith, S.~M., K.~L. Miller, G.~Salimi-Khorshidi, M.~Webster, C.~F. Beckmann,
  T.~E. Nichols, J.~D. Ramsey, and M.~W. Woolrich (2011).
\newblock Network modelling methods for f{MRI}.
\newblock {\em NeuroImage\/}~{\em 54\/}(2), 875--891.

\bibitem[\protect\citeauthoryear{Stelzl}{Stelzl}{1986}]{Stelzl_I-1986}
Stelzl, I. (1986).
\newblock Changing a causal hypothesis without changing the fit: some rules for
  generating equivalent path models.
\newblock {\em Multivariate Behavioral Research\/}~{\em 21\/}(3), 309--331.

\bibitem[\protect\citeauthoryear{Stephan, Harrison, Kiebel, David, Penny, and
  Friston}{Stephan et~al.}{2007}]{Stephan-2007}
Stephan, K.~E., L.~M. Harrison, S.~J. Kiebel, O.~David, W.~D. Penny, and K.~J.
  Friston (2007).
\newblock Dynamic causal models of neural system dynamics: current state and
  future extensions.
\newblock {\em Journal of Biosciences\/}~{\em 32}, 129--144.

\bibitem[\protect\citeauthoryear{Stephan and Roebroeck}{Stephan and
  Roebroeck}{2012}]{Stephan-2012b}
Stephan, K.~E. and A.~Roebroeck (2012).
\newblock A short history of causal modeling of f{MRI} data.
\newblock {\em NeuroImage\/}~{\em 62\/}(2), 856--863.

\bibitem[\protect\citeauthoryear{Tanner}{Tanner}{1994}]{Tanner-1994}
Tanner, M.~A. (1994).
\newblock {\em Tools for Statistical Inference -- Methods for the Exploration
  of Posterior Distributions and Likelihood Functions\/} (2nd ed.).
\newblock Springer Series in Statistics. Springer, New York.

\bibitem[\protect\citeauthoryear{Valdes-Sosa, Roebroeck, Daunizeau, and
  Friston}{Valdes-Sosa et~al.}{2011}]{Valdes-Sosa-2011}
Valdes-Sosa, P.~A., A.~Roebroeck, J.~Daunizeau, and K.~Friston (2011).
\newblock Effective connectivity: Influence, causality and biophysical
  modeling.
\newblock {\em NeuroImage\/}~{\em 58\/}(2), 339--361.

\bibitem[\protect\citeauthoryear{Verma and Pearl}{Verma and
  Pearl}{1990}]{Verma-1990b}
Verma, T.~S. and J.~Pearl (1990).
\newblock Equivalence and synthesis of causal models.
\newblock In P.~P. Bonissone, M.~Henrion, L.~N. Kanal, and J.~F. Lemmer (Eds.),
  {\em 6th Annual Conference on Uncertainty in Artificial Intelligence
  (UAI-90)}, Volume~6, pp.\  255--268. Elsevier, Amsterdam, NL.

\end{thebibliography}

\end{document}